\documentclass[prx,twocolumn,superscriptaddress,floatfix,nofootinbib,aps]{revtex4-2}

\usepackage{graphicx}% Include figure files
\usepackage{dcolumn}% Align table columns on decimal point
\usepackage{bm}% bold math

\usepackage[utf8]{inputenc}
\usepackage[T1]{fontenc}
\usepackage{booktabs,array,float,tabularx,lipsum,amsmath,multirow}
\usepackage{amssymb}
\usepackage{siunitx,xcolor,comment}
\usepackage[version=4]{mhchem}
\graphicspath{{s/}{figs/}} % 读取图片从figs/文件夹中读取，而不是当前文件夹

\usepackage[colorlinks,linkcolor=blue,anchorcolor=blue,citecolor=blue]{hyperref}
\usepackage{newtxtext,newtxmath}
\usepackage{braket} 
\usepackage{quantikz}
\usepackage{caption}
\usepackage{subcaption}
\makeatletter
\long\def\@makecaption#1#2{%
  \par\addvspace{\abovecaptionskip}%
  \begingroup\small\leftskip=0pt\rightskip=0pt\parfillskip=0pt plus 1fil\relax
  \noindent#1\ #2\par\endgroup
  \addvspace{\belowcaptionskip}}
\makeatother

\begin{document}

% 若要在预印本模式下将您所在机构的报告编号置于标题页的右上角，请使用 \preprint 命令。
% 允许使用多个 \preprint 命令。
% 如有必要，可使用 'preprintnumbers' 类选项来覆盖期刊默认设置以显示编号
%\preprint{}

\title{Optimized EIT-Based Multi-Target $\mathrm{CNOT}^{k}$ Gates in\\Heteronuclear Rydberg Atom Arrays}

% 如有必要，请重复 \author .. \affiliation 等内容
% \email、\thanks、\homepage、\altaffiliation 均适用于当前作者。解释性文字应放在 [] 中，实际电子邮件地址或网址应放在 {} 中，用于 \email 和 \homepage。请为每种类型的信息使用相应的宏。
% \affiliation 命令适用于自上次使用 \affiliation 命令以来的所有作者。该命令应置于其他信息之后。
% \affiliation 命令之后还可以跟 \email、\homepage、\thanks 等。
%\author{}
%\thanks{}
%\affiliation{}

\author{Zeyu~Zhou}
\affiliation{School of Physics, Beihang University, Beijing 100191, China}

\author{Xian-Lei~Sheng}
\email{xlsheng@buaa.edu.cn}
\affiliation{School of Physics, Beihang University, Beijing 100191, China}

\author{Peng Xu}
\email{etherxp@wipm.ac.cn}
\affiliation{State Key Laboratory of Magnetic Resonance and Atomic and Molecular Physics, Wuhan Institute of Physics and Mathematics, Innovation Academy for Precision Measurement Science and Technology, Chinese Academy of Sciences, Wuhan 430071, China}

\author{Jian Cui}
\email{jiancui@buaa.edu.cn}
\affiliation{School of Physics, Beihang University, Beijing 100191, China} 

%如需协作名称（需要在 \documentclass 中使用 superscriptaddress 选项）。必须使用 \noaffiliation（也可与 \author 命令一起使用）。
%\collaboration 可以后跟 \email、\homepage、\thanks 等。

\date{\today}

\begin{abstract}
Efficient stabilizer readout requiring multi-qubit coupling is a core bottleneck for quantum error correction. One feasible method is direct implementation of the controlled-U gate between one ancilla qubit and the data qubits assigned to stabilizer U measurements. We systematically analyze the native multi-target $\mathrm{C}^1\mathrm{NOT}^k$ gates proposed by M\"uller \emph{et al.} [Phys. Rev. Lett. 102, 170502 (2009)], which is realized via electromagnetically induced transparency (EIT) and Rydberg blockade mechanisms. Using a microscopic open-system model, we analyze the gate’s scaling with target number k and identify spontaneous emission, Doppler dephasing, target atom inter-coupling, and technical noise as major error contributions. We further optimize the protocol combining two-photon STIRAP control, heteronuclear interaction engineering, and waveform optimization. Our optimized heteronuclear protocol reaches fidelities of $98.03\%$ ($\mathrm{C}^1\mathrm{NOT}^{1}$) and $96.54\%$ ($\mathrm{C}^1\mathrm{NOT}^{4}$), in the presence of all primary noise sources and realistic experimental parameters. These results demonstrate that EIT-based multi-target gates serve as a practical building block for low-depth stabilizer readout.
\end{abstract}

%\keywords{}

\maketitle

%\section{}
% 在 \section 的参数中放置 \label 以实现交叉引用
%\subsection{}
%\subsubsection{}

% 若处于双栏模式，此环境将切换为单栏格式，以便展示长方程。请谨慎使用。
%\begin{widetext}
% 在此处输入长方程
%\end{widetext}

% 图形应作为浮动体放入文本中。
% 使用 graphics 或 graphicx 包（随 LaTeX2e 分发）
% 并使用这些包中定义的 \includegraphics 宏。
% 这是图形的一般形式示例：
% 在 \caption{} 命令的花括号中填写说明文字。在 \label{} 命令的花括号中填入您将用于 \ref{} 命令的标签。
% 如果图形应横跨整个页面，请使用 figure* 环境。无需进行显式居中设置。
% \begin{figure}
% \end{figure}
% 将图形环境用转页环境包围以实现横向布局
% 图形
% \begin{figure}
% \end{figure}

% 表格应作为浮动对象出现在文本中%
% 这是一个表格的一般形式示例：
% 在 \caption{} 命令的花括号中填写标题。在 \label{} 命令的花括号中填入您将用于 \ref{} 命令的标签。
% 在 \begin{tabular}{} 命令的空花括号中插入列说明符（l、r、c、d 等）。
% ruledtabular 环境为表格添加双线，并设置合理的默认表格设置。
% 使用 table* 环境在双栏排版中获得全宽表格。
% 添加 \usepackage{longtable} 并使用 longtable（或 longtable*）环境来获得格式优美的长表格。或者使用 [H] 放置选项来断开长表格（但不如 longtable 控制精细）。
% \begin{table}%[H] 添加 [H] 放置选项以使表格跨页断开
% \begin{ruledtabular}
% 表格内容行，每行以 \\ 结尾
% \end{tabular}
% \end{ruledtabular}
% \end{table}
% 用turnpage环境将表格环境包围起来以实现横向排版
% 表格
% \begin{table}
% \begin{ruledtabular}
% \end{tabular}
% \end{ruledtabular}
% \end{table}

\section{Introduction} \label{sec:intro}
Quantum computing has advanced rapidly, but a substantial gap remains between current hardware and large-scale, practically useful quantum computation.
Closing this gap requires more than increasing qubit number: it also requires high-fidelity quantum operations, long coherence times, and architectures that support repeated error correction.

Neutral atoms are widely regarded as a promising platform for scalable quantum computing because they combine weak environmental coupling, long coherence times, and intrinsic scalability~\cite{neutral-atoms1,neutral-atoms2}.
Advances in optical lattices and optical tweezers have enabled precise trapping and coherent manipulation of individual ultracold atoms~\cite{control-of-qubit1,control-of-qubit2,control-of-qubit3}, as well as the assembly of defect-free arrays with programmable geometries~\cite{bernien_51_atom_2017,2d-array2,2d-array3}.
Together with strong and controllable Rydberg interactions, these capabilities have enabled programmable gate-based neutral-atom processors~\cite{integrability}.
Recent experiments have further demonstrated architectural progress, including optical tweezer arrays with more than 6,100 highly coherent atomic qubits~\cite{tweezer-array}, the trapping of 11,000 individual atoms in a metasurface-generated tweezer array~\cite{chen_11000_2026}, and key ingredients of fault-tolerant neutral-atom processing~\cite{QEC3}.

The central challenge is therefore no longer qubit number alone, but accurate quantum logic under realistic experimental conditions.
For Rydberg-based neutral-atom gates, practical performance is often limited by spontaneous emission, Doppler dephasing, laser noise, detuning fluctuations, and crosstalk~\cite{imperfection1,laser-phase-noise1,laser-phase-noise2,doppler-dephasing1,doppler-dephasing2,warttmann_suppressing_2026}.
Although neutral-atom CZ and CNOT gates have seen important experimental progress~\cite{CZ-neutral1,CZ-neutral2,CNOT-neutral1,CNOT-neutral2,CNOT-neutral3}, reliable error correction also requires gate protocols whose error mechanisms remain controlled when operations are repeated many times.

Quantum error correction (QEC) is essential for scalable quantum computation~\cite{QEC1,QEC2,QEC3}.
Its effectiveness depends strongly on the structure and quality of the underlying physical gates.
If stabilizer measurements must be decomposed into long sequences of two-qubit gates, circuit depth and error accumulation can severely limit performance.
Native multi-qubit gates are therefore attractive because they can reduce circuit depth and simplify syndrome extraction~\cite{QEC2,jandura_surface_2026,locher_multiqubit_2026}.
In particular, if a weight-$k$ stabilizer can be implemented directly as a native $k$-body operation, the corresponding readout circuit can in principle be compressed from $\mathcal{O}(k)$ to $\mathcal{O}(1)$ depth.

A representative example is the toric code, whose Hamiltonian contains intrinsic four-body stabilizer terms~\cite{kitaev_fault-tolerant_2003}.
In Rydberg-atom architectures, Auger \emph{et al.} proposed a surface-code blueprint in which multi-qubit gates are engineered through electromagnetically induced transparency (EIT)~\cite{surfacecode}.
More recent theoretical and experimental work has further highlighted the relevance of dual-species arrays and multi-qubit Rydberg operations for fast stabilizer readout~\cite{petrosyan_fast_2024,experimental-setup3,wang_multi-qubit_2026}.
These developments point to a direct route toward scalable neutral-atom QEC: engineer native multi-qubit operations that match the structure of stabilizer readout.

Motivated by this perspective, we study high-fidelity multi-target controlled gates in a Rydberg atom array.
Specifically, we focus on an effective four-body operation implemented through a native $\mathrm{C}^1\mathrm{NOT}^{4}$ gate, which is relevant for toric-code-inspired stabilizer measurements.
Building on the EIT-based protocol proposed by M\"uller \emph{et al.}~\cite{EIT-protocol}, we examine three complementary refinements.
First, we replace the single-photon excitation of the control atom with a two-photon stimulated Raman adiabatic passage (STIRAP) process~\cite{STIRAP1,STIRAP2} to reduce the required coupling strength and suppress intermediate-state scattering.
Second, we employ a heteronuclear Rydberg-atom array to engineer a favorable interaction hierarchy and reduce crosstalk among target atoms~\cite{dual-species1,beterov_rydberg_2015,anand_dual-species_2024,ireland_interspecies_2024}.
Third, we integrate optimal-control techniques~\cite{optimal-control1,DCRAB1,DCRAB2,kazemi_multiqubit_2025,stein_multi-target_2025} to optimize the pulse design under realistic constraints.
We then assess an experimentally feasible implementation of a four-target $\mathrm{C}^1\mathrm{NOT}^{4}$ gate and benchmark its fidelity and robustness against the dominant dissipation channels and technical-noise sources.

The remainder of the paper is organized as follows.
Sec.~\ref{sec:model} introduces the EIT-based multi-target gate model and its Hamiltonian.
Sec.~\ref{sec:open} defines the open-system dynamics and gate-performance metrics.
Sec.~\ref{sec:dominant-errors} analyzes the dominant error mechanisms.
Sec.~\ref{sec:opt} presents the heteronuclear interaction engineering and optimal-control strategy.
Sec.~\ref{sec:experimental_mapping} maps the protocol to experimentally realistic parameters, and Sec.~\ref{sec:conclusion} summarizes the main conclusions.
%%%%%%%%%%%%%%%%
\section{From Quantum Error Correction to $\mathrm{CNOT}^{k}$ Gates}

\subsection{Four-body interactions and stabilizer readout}

Many topological and stabilizer codes are defined by weight-$k$ multi-qubit parity checks.
A representative example is the Toric Code on a square lattice, whose stabilizers are
\begin{equation}
A_s=\prod_{j\in\mathrm{star}(s)} X_j,
\qquad
B_p=\prod_{j\in\mathrm{plaquette}(p)} Z_j,
\label{eq:toric_stabilizers_paper_en}
\end{equation}
where both $A_s$ and $B_p$ correspond to four-body Pauli measurements ($k=4$). In practical
fault-tolerant architectures, however, the central primitive is not necessarily a static
many-body Hamiltonian, but the repeated extraction of stabilizer eigenvalues $\pm1$
(syndromes) while preserving the encoded quantum information in the data register, apart from the intended projection associated with the measured stabilizer. This task is naturally
formulated as an ancilla-assisted measurement of a multi-qubit Pauli operator.

Let $S$ be a weight-$k$ Pauli operator with eigenvalues $\pm1$. Preparing an ancilla
qubit in $\ket{+}_a=(\ket{0}_a+\ket{1}_a)/\sqrt{2}$ and applying the controlled-stabilizer
unitary
\begin{equation}
U_{cS}
=
\ket{0}\!\bra{0}_a\otimes I
+
\ket{1}\!\bra{1}_a\otimes S
\label{eq:controlledS_paper_en}
\end{equation}
yields, for an eigenstate $S\ket{\psi}=s\ket{\psi}$ with $s\in\{\pm1\}$,
\begin{equation}
U_{cS}\ket{+}_a\ket{\psi}
=
\frac{1}{\sqrt{2}}
\bigl(\ket{0}_a+s\ket{1}_a\bigr)\ket{\psi}.
\label{eq:kickback_paper_en}
\end{equation}
A subsequent $X$-basis measurement of the ancilla returns the stabilizer eigenvalue
$s$ while ideally preserving the data state. This is precisely the phase-kickback
mechanism underlying standard stabilizer readout.

\begin{figure}[h]
\centering
    \begin{subfigure}{0.8\linewidth}
        \centering
        \caption{}
        \includegraphics[width=\linewidth]{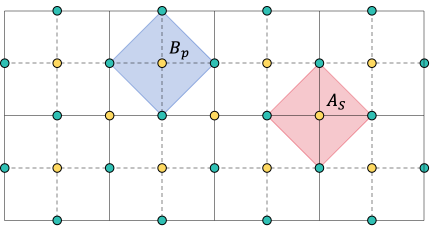}
    \end{subfigure}

    \begin{subfigure}{0.8\linewidth}
        \centering
        \caption{}
        \begin{quantikz}[row sep=0.22cm, column sep=0.35cm]
        \lstick{$\ket{0}_a$} & \gate{H} & \ctrl{1} & \ctrl{2} & \ctrl{3} & \ctrl{4} & \gate{H} & \meter{} \\
        \lstick{$q_1$}       & \qw      & \targ{}  & \qw      & \qw      & \qw      & \qw      & \qw \\
        \lstick{$q_2$}       & \qw      & \qw      & \targ{}  & \qw      & \qw      & \qw      & \qw \\
        \lstick{$q_3$}       & \qw      & \qw      & \qw      & \targ{}  & \qw      & \qw      & \qw \\
        \lstick{$q_4$}       & \qw      & \qw      & \qw      & \qw      & \targ{}  & \qw      & \qw
        \end{quantikz}
    \end{subfigure}
\caption{From four-body stabilizers to ancilla-assisted syndrome extraction. (a) The Toric Code on a square lattice is defined by weight-4 star and plaquette stabilizers. Yellow circles denote ancillary qubits, and green circles denote data qubits. (b) Standard circuit for measuring an $X$-type four-body stabilizer using an ancillary qubit and four CNOT gates. A native shared-control multi-target gate may compress this sequential entangling layer into a constant-depth block.}
\label{fig:stabilizer_readout_paper_en}
\end{figure}

Fig.~\ref{fig:stabilizer_readout_paper_en}(a) shows the square-lattice geometry of the
Toric Code, where the star operator $A_s$ and plaquette operator $B_p$ are both
weight-4 stabilizers. Fig.~\ref{fig:stabilizer_readout_paper_en}(b) illustrates the
standard ancilla-assisted circuit for measuring an $X$-type four-body stabilizer:
the ancilla is prepared in the $X$ basis, interacts sequentially with the four data
qubits through four CNOT gates, and is then measured again in the $X$ basis.

Equation~\eqref{eq:controlledS_paper_en} is directly implementable when the platform
provides a shared-control multi-target entangling primitive. For a $Z$-type four-body
stabilizer $B_p=Z_1Z_2Z_3Z_4$, one may realize $U_{cS}$ via a multi-target controlled-phase
operation,
\begin{align}
U_{cB_p}
&=
\prod_{j=1}^{4}\mathrm{CZ}_{a,j}
\nonumber\\
&=
\ket{0}\!\bra{0}_a\otimes I^{\otimes4}
+
\ket{1}\!\bra{1}_a\otimes (Z_1Z_2Z_3Z_4).
\label{eq:cz4_paper_en}
\end{align}
Similarly, for an $X$-type four-body stabilizer $A_s=X_1X_2X_3X_4$, a shared-control
multi-target $\mathrm{CNOT}$ realizes the corresponding controlled-stabilizer operation,
\begin{align}
U_{cA_s}
&=
\prod_{j=1}^{4}\mathrm{CNOT}_{a,j}
\nonumber\\
&=
\ket{0}\!\bra{0}_a\otimes I^{\otimes4}
+
\ket{1}\!\bra{1}_a\otimes (X_1X_2X_3X_4).
\label{eq:cnot4_paper_en}
\end{align}
Consequently, if the hardware supports a native shared-control multi-target gate, then
the entangling part of a weight-$k$ stabilizer readout can be compressed from
$\mathcal{O}(k)$ sequential two-qubit operations into an $\mathcal{O}(1)$-depth block.
This reduction is particularly attractive for repeated syndrome extraction in QEC.

\subsection{$\mathrm{CNOT}^{k}$ as a scalable primitive}

Motivated by the constant-depth implementation of weight-$k$ stabilizer readout, we
consider the one-control, $k$-target $\mathrm{CNOT}^{k}$ gate defined by
\begin{equation}
\ket{c}\ket{t_1\cdots t_k}
\mapsto
\ket{c}\ket{t_1\oplus c,\ldots,t_k\oplus c},
\qquad
c,t_j\in\{0,1\}.
\label{eq:cnotk_paper_en}
\end{equation}
It is locally equivalent to the multi-target controlled-$Z$ operation
\begin{equation}
\mathrm{CZ}^{k}
=
\prod_{j=1}^{k}\mathrm{CZ}_{c,j}
=
\ket{0}\!\bra{0}_c\otimes I^{\otimes k}
+
\ket{1}\!\bra{1}_c\otimes (Z_1Z_2\cdots Z_k),
\label{eq:czk_paper_en}
\end{equation}
via Hadamard gates on each target qubit. Therefore, engineering high-fidelity
$\mathrm{CNOT}^{k}$ gates provides a direct route toward shared-control multi-qubit
entangling primitives required in stabilizer measurement.

From the standpoint of QEC, the stabilizer readout layer is executed repeatedly and
often dominates the spacetime overhead. As a consequence, the physical fidelity and
noise structure of $\mathrm{CNOT}^{k}$ gates directly determine the quality of the
extracted syndrome information and, ultimately, the logical performance. For a native
implementation executed in a single entangling block of duration $T_g$, a generic
error-budget decomposition can be written as
\begin{equation}
1-\mathcal{F}(k)
\approx
\epsilon_{\mathrm{ctrl}}
+
k\,\epsilon_{\mathrm{tgt}}
+
\epsilon_{\mathrm{corr}}(k),
\label{eq:scaling_paper_en}
\end{equation}
where $\epsilon_{\mathrm{ctrl}}$ captures errors associated with the control atom that are approximately
independent of $k$, $k\,\epsilon_{\mathrm{tgt}}$ accounts for dissipation local to each target atom
and technical noise accumulated during $T_g$, and $\epsilon_{\mathrm{corr}}(k)$ denotes
correlated contributions that become increasingly relevant as $k$ grows, such as
imperfect blockade, residual interactions between target atoms, crosstalk, and spatial
inhomogeneity. A scalable design therefore aims to keep $T_g$ nearly independent of
$k$ while suppressing $\epsilon_{\mathrm{corr}}(k)$, so that the total error remains
dominated by controllable single-particle channels.

These considerations motivate gate protocols that minimize intermediate-state
population and reduce sensitivity to Doppler dephasing and laser noise, while
simultaneously engineering an interaction pattern that clearly separates the intended
interaction between the control atom and each target atom from unwanted interactions between target atoms. In the following, we
focus on EIT-based multi-target gates as a systematic route toward high-fidelity and
scalable $\mathrm{CNOT}^{k}$ operations.
%%%%%%%%%%%%%%%%
\section{Model and Hamiltonian} \label{sec:model}

This section specifies the model and Hamiltonian for the EIT-based Rydberg gate.
We first summarize the EIT mechanism that underlies the native
$\mathrm{C}^1\mathrm{NOT}^1$ gate proposed by M\"uller \emph{et al.}~\cite{EIT-protocol}.
We then extend the model to the $\mathrm{C}^1\mathrm{NOT}^k$ gate and identify
the new scalability bottleneck: interactions between target atoms generate an
interaction-induced effective detuning that spoils the dark-state protection.

\begin{figure}[t]
   \centering
   \includegraphics[width=0.38\textwidth]{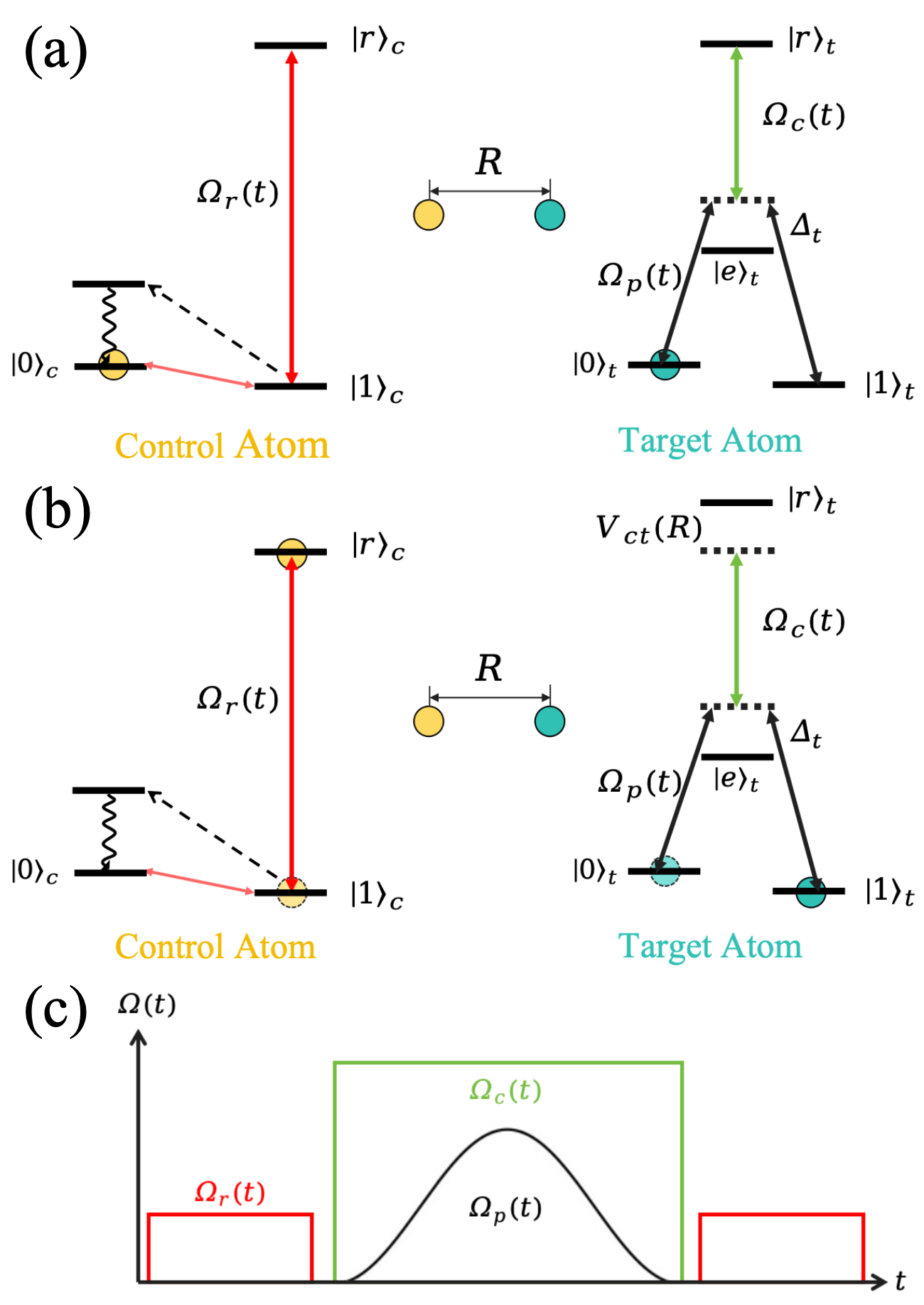}
\caption{EIT-based CNOT protocol proposed by M\"uller \emph{et al.}~\cite{EIT-protocol}. (a) Energy-level structure when the control and target atoms are initially in $|0\rangle_c$ and $|0\rangle_t$, respectively, with separation $R$. (b) Conditional branch for initial state $|1\rangle_c|0\rangle_t$. A $\pi$ pulse excites the control atom to $|r\rangle_c$, and the resulting Rydberg blockade breaks the target EIT resonance, producing the conditional transformation $|0\rangle_t\rightarrow |1\rangle_t$. (c) Pulse sequence for $\Omega_p(t)$, $\Omega_c(t)$, and $\Omega_r(t)$. The control pulse satisfies $\int\Omega_r(t)dt=\pi$, while the target adiabatic pulse satisfies $\int \Omega_p^2(t)dt/\Delta=\pi$.}
      \label{fig:EIT-protocol}
\end{figure}

\subsection{EIT gate protocol}\label{subsec:eit_mech}

As shown in Fig.~\ref{fig:EIT-protocol}, we consider a control atom (subscript $c$) and a target atom (subscript $t$).
The control atom is driven on the ground--Rydberg transition
$|1\rangle_c\leftrightarrow|r\rangle_c$ by a pulse $\Omega_r(t)$.
The target atom is driven in a $\Lambda$-type EIT configuration.
Both qubit states $|0\rangle_t$ and $|1\rangle_t$ couple to an intermediate
excited state $|e\rangle_t$ via a probe field $\Omega_p(t)$, while
$|e\rangle_t$ couples to the Rydberg state $|r\rangle_t$ via a coupling field
$\Omega_c(t)$. In the rotating frame,
\begin{align}
\mathcal{H}_{c}(t)
&=\frac{\hbar\Omega_r(t)}{2}\Bigl(|1\rangle_c\langle r|+|r\rangle_c\langle 1|\Bigr)
-\hbar\delta_c\,|r\rangle_c\langle r| ,
\label{eq:Hc}\\
\mathcal{H}_{t}(t)
&=\frac{\hbar\Omega_p(t)}{2}\Bigl(|0\rangle_t\langle e|+|1\rangle_t\langle e|\Bigr)
+\frac{\hbar\Omega_c(t)}{2}\,|e\rangle_t\langle r|+\mathrm{h.c.} \nonumber\\
&\qquad-\hbar\Delta_t\,|e\rangle_t\langle e|
-\hbar\delta_t\,|r\rangle_t\langle r| .
\label{eq:Ht}
\end{align}
When both atoms occupy Rydberg states, the van der Waals interaction induces a
diagonal energy shift
\begin{align}
\mathcal{H}_{\mathrm{int}}^{ct}
=V_{ct}(R)\,\hat n_{rc}\otimes \hat n_{rt},
\quad
\hat n_{rc}=|r\rangle_c\langle r|,
\;
\hat n_{rt}=|r\rangle_t\langle r|,
\label{eq:Hct}
\end{align}
where $R$ is the separation between the control atom and the target atom. The total Hamiltonian for the
two-atom problem is
\begin{align}
\mathcal{H}_{\mathrm{C}^1\mathrm{NOT}^1}(t)
=\mathcal{H}_c(t)
+ \mathcal{H}_t(t)
+\mathcal{H}_{\mathrm{int}}^{ct}.
\label{eq:HCNOT1}
\end{align}
To expose the dark-state structure, we introduce the $\sigma_x$ eigenbasis of the target atom,
\begin{align}
|d_1\rangle_t=|-\rangle_t=\frac{|1\rangle_t-|0\rangle_t}{\sqrt{2}},
\qquad
|+\rangle_t=\frac{|1\rangle_t+|0\rangle_t}{\sqrt{2}} .
\label{eq:plus_minus}
\end{align}
Only $|+\rangle_t$ couples to $|e\rangle_t$ via $\Omega_p(t)$, while
$|d_1\rangle_t$ is decoupled.
In the subspace spanned by $\{|+\rangle_t,|e\rangle_t,|r\rangle_t\}$, when the
intermediate-state detuning is large,
\begin{align}
|\Delta_t|\gg|\Omega_p(t)|,|\Omega_c(t)|,
\label{eq:adiabatic_elim_cond}
\end{align}
the intermediate state $|e\rangle_t$ can be adiabatically eliminated.
The driven system then admits an approximate dark eigenstate with suppressed
$|e\rangle_t$ population.
When the effective two-photon detuning is near zero, the instantaneous dark state
can be written as~\cite{EIT-protocol}
\begin{align}
|d_2(t)\rangle_t
=\frac{1}{\sqrt{1+x^2(t)}}\Bigl(|+\rangle_t-x(t)\,|r\rangle_t\Bigr),
\label{eq:dark_state}
\end{align}
where $x(t)=\sqrt{2}\,\Omega_p(t)/\Omega_c(t)$.
A detailed derivation of the dark-state picture and the effective-model reduction is provided in Appendix~\ref{app:eit_dynamics}.

The essential conditionality is that the control-atom Rydberg excitation shifts the
target-atom Rydberg level via Eq.~\eqref{eq:Hct}.
This shift can be viewed as an interaction-induced modification of the target-atom
two-photon detuning during the EIT stage,
\begin{align}
\delta_t^{(\mathrm{eff})}
=\delta_t-\frac{V_{ct}(R)}{\hbar}\,n_{rc},
\qquad n_{rc}\in\{0,1\}.
\label{eq:delta_eff_control}
\end{align}
If the control atom remains in $|0\rangle_c$ ($n_{rc}=0$), each target atom adiabatically
follows the dark state in Eq.~\eqref{eq:dark_state} and returns to its qubit
manifold with negligible scattering from $|e\rangle_t$.
If the control atom is promoted to $|r\rangle_c$ ($n_{rc}=1$), the induced shift makes
$\delta_t^{(\mathrm{eff})}$ appreciable and spoils the dark-state protection.
The target atom then undergoes a nontrivial population transfer that implements the
desired conditional transformation.
Combined with the standard $\pi$--$2\pi$--$\pi$ pulse sequence~\cite{EIT-protocol},
this realizes a native $\mathrm{CNOT}$ gate on the computational basis, up to
single-qubit phases that can be compensated.

\subsection{Extension to $\mathrm{C}^1\mathrm{NOT}^k$}\label{subsec:ext_k}
We now generalize to one control atom and $k$ target atoms driven in parallel by
the same EIT pulses. The total Hamiltonian reads
\begin{align}
\mathcal{H}_{\mathrm{C}^1\mathrm{NOT}^k}(t)
&=\mathcal{H}_c(t)
+\sum_{i=1}^{k} \mathcal{H}_{t_i}(t)
+\mathcal{H}_{\mathrm{int}}^{ct}(t)
+\mathcal{H}_{\mathrm{int}}^{tt}(t),
\label{eq:HCNOTk_total}
\end{align}
where $\mathcal{H}_{t_i}(t)$ is given by \eqref{eq:Ht} acting on target $i$. The interaction terms are
\begin{align}
\mathcal{H}_{\mathrm{int}}^{ct}(t)
&=\sum_{i=1}^{k} V_{ct_i}^{(i)}(R_{ct_i})\,\hat n_{rc}\otimes \hat n_{rt_i},
\label{eq:Hct_many}\\
\mathcal{H}_{\mathrm{int}}^{tt}(t)
&=\sum_{1\le i<j\le k} V_{t_it_j}(R_{t_it_j})\,\hat n_{rt_i}\hat n_{rt_j},
\label{eq:Htt_many}
\end{align}
with $\hat n_{rt_i}=|r\rangle_{t_i}\langle r|$ the Rydberg projector of target $i$.

In the single-target case, when $n_{rc}=0$, the target atom can follow the dark state
and remain protected.
For $k>1$, however, each target atom acquires a small but finite transient Rydberg
population during adiabatic evolution.
As a result, the interaction between target atoms in Eq.~\eqref{eq:Htt_many} produces
additional energy shifts that act as an interaction-induced detuning, a mechanism
absent in the original single-target analysis.
At a mean-field level, the effective two-photon detuning for target $i$ can be
written as
\begin{align}
\delta_{t_i}^{(\mathrm{eff})}(t)
\simeq \delta_t
-\frac{V_{ct_i}(R_{ct_i})}{\hbar}\,n_{rc}
-\frac{1}{\hbar}\sum_{j\ne i} V_{t_it_j}(R_{t_it_j})\,\langle \hat n_{rt_j}\rangle.
\label{eq:delta_eff_k}
\end{align}
Even when $n_{rc}=0$, the last term is generally nonzero, deteriorating the dark-state condition and inducing systematic errors as $k$ increases.

\iffalse
\subsubsection{Error trends versus $k$ and dissipation strength}
To incorporate dissipation and dephasing, we use a Lindblad master equation
\begin{align}
\dot\rho
=-\frac{i}{\hbar}\bigl[\mathcal{H}_{\mathrm{C}^1\mathrm{NOT}^k}(t),\rho\bigr]
+\sum_\mu\left(L_\mu\rho L_\mu^\dagger-\frac{1}{2}\{L_\mu^\dagger L_\mu,\rho\}\right),
\label{eq:master_eq}
\end{align}
where $L_\mu$ account for intermediate-state spontaneous decay, Rydberg decay, and dephasing on the ground--Rydberg coherence. Since EIT suppresses the intermediate-state population, the dominant errors originate from (i) imperfect adiabaticity and (ii) transient Rydberg admixture, whose impact grows with the interaction-induced detuning in \eqref{eq:delta_eff_k}. Consequently, the gate infidelity typically increases with $k$ because the number of interacting target-atom pairs scales as $k(k-1)/2$, and it is further amplified as the dissipation strengths increase. In the following sections, we quantify these trends numerically.
\fi

%%%%%%%%%%%%%%%%%%%%%%%%%%%%%%%
\section{Open-System Dynamics}
\label{sec:open}

In realistic experiments, the $\mathrm{C}^1\mathrm{NOT}^k$ protocol is implemented
in an open quantum system and is therefore subject to dissipation and dephasing.
In this section, we specify the Lindblad master equation together with the noise
channels (jump operators) used in our simulations, and define the
gate-performance metrics employed throughout this work.

\subsection{Lindblad master equation and noise channels}
\label{subsec:lindblad}

\subsubsection{Master equation}
We describe the density matrix $\rho(t)$ of the full $\mathrm{C}^1\mathrm{NOT}^k$
system by the Lindblad master equation
\begin{align}
\dot{\rho}(t)
&=
-\frac{i}{\hbar}\bigl[\mathcal{H}(t),\rho(t)\bigr]
\nonumber\\
&\quad+\sum_{\mu}\left(
L_{\mu}\rho(t)L_{\mu}^{\dagger}
-\frac{1}{2}\left\{L_{\mu}^{\dagger}L_{\mu},\rho(t)\right\}
\right),
\label{eq:lindblad_master}
\end{align}
where $\mathcal{H}(t)=\mathcal{H}_{\mathrm{C}^1\mathrm{NOT}^k}(t)$ is the many-body
Hamiltonian defined in Sec.~\ref{sec:model}. The index $\mu$ labels all jump
operators, including different atoms and different noise processes.

\subsubsection{Spontaneous emission}
In the ideal coherent EIT/STIRAP limit, destructive interference in the
dark state suppresses intermediate-state population even on single-photon
resonance.  In a realistic open system, however, spontaneous emission,
dephasing, and imperfect adiabatic following perturb this dark-state
evolution and generate a small transient population in the intermediate
excited states~\cite{STIRAP1}.  Because their natural lifetimes are
comparable to typical gate durations, the resulting scattering cannot be
neglected.

\begin{figure}[t]
   \centering
   \includegraphics[width=0.38\textwidth]{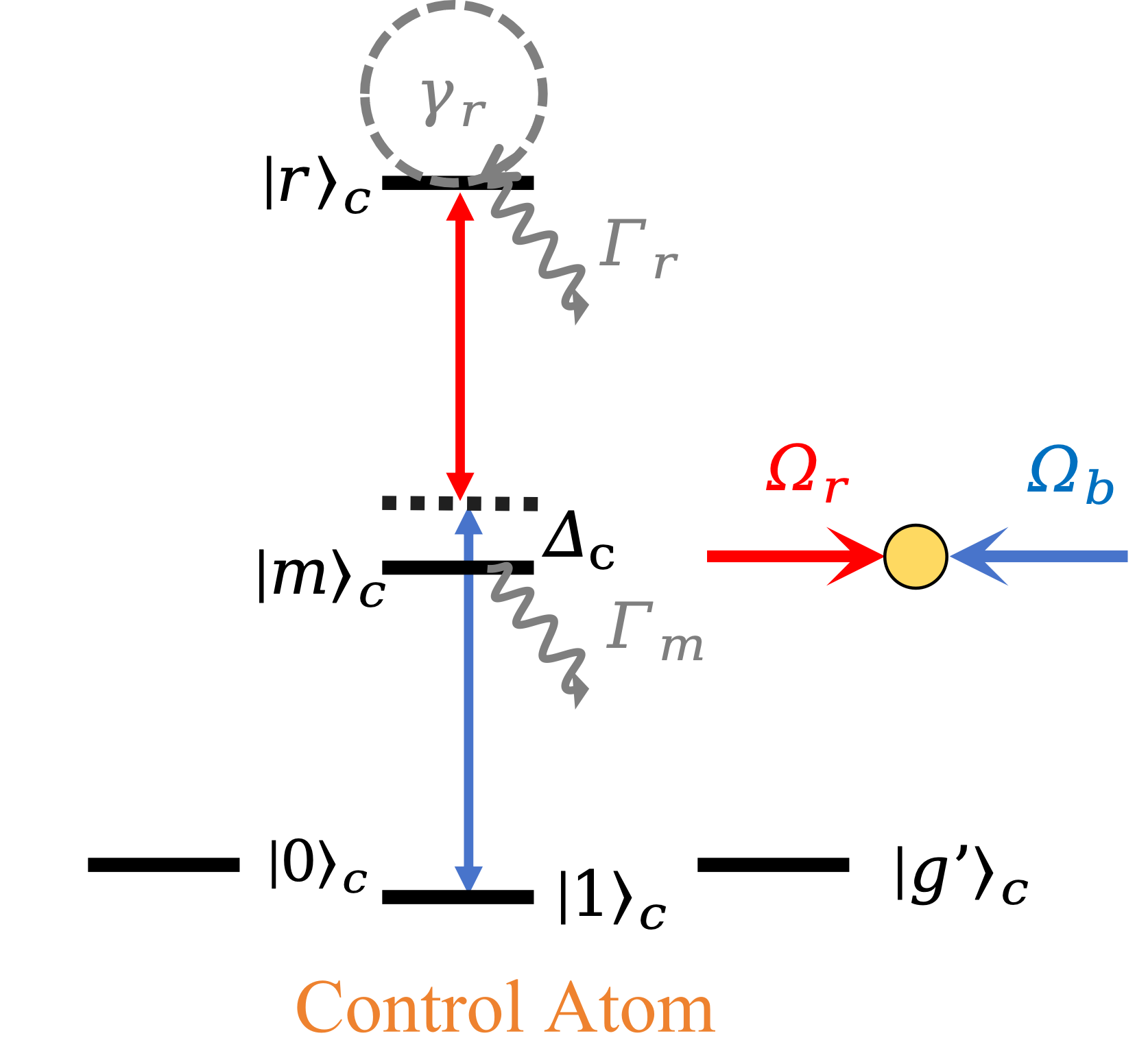}
   \caption{Five-level model for the control atom used in the
   Lindblad simulation. The computational states $|0\rangle_c$ and
   $|1\rangle_c$ are coupled to the intermediate state $|m\rangle_c$ and
   the Rydberg state $|r\rangle_c$ by the two STIRAP fields
   $\Omega_b$ and $\Omega_r$. The intermediate state decays at rate
   $\Gamma_m$ (denoted by $\Gamma$ in the Lindblad model) into the computational manifold and the loss state
   $|g'\rangle_c$, while $\gamma_r$ denotes Rydberg-state dephasing.
   The target-atom model has the same dissipative structure, with
   $|m\rangle_c$ replaced by $|e\rangle_t$.}
   \label{fig:control_lindblad_levels}
\end{figure}

To model this in a minimal yet practical way, we adopt a five-level description
for each atom,
$\{|0\rangle,|1\rangle,|e/m\rangle,|r\rangle,|g'\rangle\}$.
The level structure and the associated dissipative channels are illustrated
for the control atom in Fig.~\ref{fig:control_lindblad_levels}.

The additional ground state $|g'\rangle$ collects all other ground-state
sublevels outside the computational basis.
We assume that the intermediate state decays into $|0\rangle$, $|1\rangle$, and
$|g'\rangle$ with equal branching ratios (a standard coarse-grained
approximation). Denoting the total intermediate-state decay rate by $\Gamma$,
the partial rate is $\Gamma'=\Gamma/3$, and the jump operators are
\begin{align}
L^{(a)}_{s,\alpha}
=
\sqrt{\Gamma'}\,|\alpha\rangle_a\langle \xi_a|,
\quad
\alpha\in\{0,1,g'\},
\;
a\in\{c,t_1,\dots,t_k\},
\label{eq:jump_sp}
\end{align}
where $\xi_c\equiv m$ for the control atom and $\xi_{t_i}\equiv e$ for the $i$th
target atom. In Eq.~\eqref{eq:lindblad_master}, the total spontaneous-emission
channel is obtained by summing over $a\in\{c,t_1,\dots,t_k\}$ and
$\alpha\in\{0,1,g'\}$ in Eq.~\eqref{eq:jump_sp}.

In our parameter regime, the Rydberg-state lifetime is typically orders of
magnitude longer than the gate duration; hence Rydberg spontaneous decay is
neglected at the level of accuracy targeted here.

\subsubsection{Doppler dephasing}
Finite atomic temperature induces Doppler shifts, which appear as random phase
fluctuations in laser-driven coherence. While a microscopic treatment would
sample random velocities and average over trajectories, we adopt an effective
pure-dephasing model that captures the dominant loss of ground--Rydberg
coherence:
\begin{align}
L_{d,r}^{(a)}=\sqrt{\frac{\gamma}{2}}\,|r\rangle_a\langle r|,
\quad
a\in\{c,t_1,\dots,t_k\},
\label{eq:doppler_jump}
\end{align}
where $\gamma$ is the pure-dephasing rate. This model is consistent with
interpreting Doppler motion as an effective random optical-phase detuning and
provides a convenient one-parameter characterization of dephasing strength.

\subsection{Gate-performance metrics}
\label{subsec:metrics}

\subsubsection{Average gate fidelity}
Let $\mathcal{E}$ denote the effective quantum channel acting on the
computational subspace, where leakage to non-computational states is treated as
an error process, and let $\mathcal{U}$ denote the ideal
$\mathrm{C}^1\mathrm{NOT}^k$ unitary. The performance of the gate is quantified
by the average gate fidelity
\begin{equation}
F_{\mathrm{avg}}(\mathcal{E},\mathcal{U})
=
\int d\psi\;
\langle \psi |\, \mathcal{U}^{\dagger}\circ\mathcal{E}
\bigl(|\psi\rangle\langle\psi|\bigr)\circ\mathcal{U}\, |\psi\rangle,
\label{eq:favg_def}
\end{equation}
where the integral is taken over the Haar measure on the $d$-dimensional
computational Hilbert space with $d=2^{k+1}$.

In practice, we relate $F_{\mathrm{avg}}$ to the entanglement fidelity $F_e$,
\begin{equation}
F_{\mathrm{avg}}=\frac{d\,F_e+1}{d+1},
\quad
F_e=
\langle \Phi_d |\, (\mathcal{I}\otimes \mathcal{U}^{\dagger}\circ\mathcal{E})
\bigl(|\Phi_d\rangle\langle\Phi_d|\bigr)\, |\Phi_d\rangle,
\label{eq:favg_fe}
\end{equation}
with $|\Phi_d\rangle=\frac{1}{\sqrt{d}}\sum_{j=0}^{d-1}|j\rangle\otimes|j\rangle$
the maximally entangled state and $\mathcal{I}$ the identity channel on the
reference system.

For numerical simulations, we evaluate the fidelity via the equivalent
Choi--Jamiolkowski expression. The (unnormalized) Choi matrix is
\begin{equation}
J(\mathcal{E})
=
\sum_{i,j=0}^{d-1}
\mathcal{E}\!\left(|i\rangle\langle j|\right)\otimes |i\rangle\langle j|,
\end{equation}
and for a unitary target $\mathcal{U}$ one has
\begin{equation}
F_{\mathrm{avg}}(\mathcal{E},\mathcal{U})
=
\frac{\mathrm{Tr}\!\left[
J(\mathcal{U})^{\dagger} J(\mathcal{E})
\right] + d}{d(d+1)}.
\label{eq:favg_choi}
\end{equation}
Although Eq.~\eqref{eq:favg_fe} suggests that $F_e$ may be obtained from a
single application of the channel to $|\Phi_d\rangle$, this would require
simulating an enlarged $d^2$-dimensional Hilbert space and carefully tracking
leakage. For our open-system simulations, it is numerically more convenient to
construct $\mathcal{E}$ on the computational basis and then apply
Eq.~\eqref{eq:favg_choi}.

\subsubsection{Bell-state fidelity}
As a complementary and experimentally relevant diagnostic, we also report the
Bell-state fidelity for the control entangled with the $k$-qubit target
register. We prepare the input state
\begin{align}
|\Psi_{\mathrm{in}}\rangle
=
\frac{1}{\sqrt{2}}\Bigl(
|0\rangle_c\otimes|+\rangle^{\otimes k}
+
|1\rangle_c\otimes|+\rangle^{\otimes k}
\Bigr),
\label{eq:bell_input}
\end{align}
define the ideal output $|\Psi_{\mathrm{id}}\rangle=\mathcal{U}|\Psi_{\mathrm{in}}\rangle$,
and compute
\begin{align}
F_{\mathrm{Bell}}
=
\langle \Psi_{\mathrm{id}}|\,\rho_{\mathrm{out}}\,|\Psi_{\mathrm{id}}\rangle,
\qquad
\rho_{\mathrm{out}}=\mathcal{E}\bigl(|\Psi_{\mathrm{in}}\rangle\langle\Psi_{\mathrm{in}}|\bigr).
\label{eq:fbell_def}
\end{align}
This metric is sensitive to both coherent errors and incoherent loss,
including leakage to $|g'\rangle$.

\subsection{Dimensionless dissipation parameters}
\label{subsec:dimensionless}

To make the error analysis transferable across different choices of pulse
amplitudes and gate durations, we introduce dimensionless dissipation
parameters. Let $T_g$ denote the total gate duration. We define
\begin{align}
\tilde{\Gamma}=\Gamma T_g,\qquad
\tilde{\gamma}=\gamma T_g,
\label{eq:dimensionless_rates}
\end{align}
and sweep $(\tilde{\Gamma},\tilde{\gamma})$ while keeping the pulse \emph{shapes}
fixed. Equivalently, one may normalize by a characteristic Rabi scale $\Omega_0$
(e.g., $\Omega_c^{\max}$) and use $\Gamma/\Omega_0$ and $\gamma/\Omega_0$; both
parameterizations reduce the dynamics to dimensionless control parameters.
The dependence of gate fidelity on $(\tilde{\Gamma},\tilde{\gamma})$ and on the
number of targets $k$ is analyzed in Sec.~\ref{sec:dominant-errors}.

\section{Dominant Error Sources}
\label{sec:dominant-errors}

Based on the open-system model and noise channels introduced in
Sec.~\ref{sec:open}, we now analyze the dominant physical mechanisms that
limit the performance of the $\mathrm{C}^1\mathrm{NOT}^k$ gate.
Rather than repeating the microscopic definitions of all noise channels,
this section emphasizes the physical origin of gate infidelity and its
scaling with the number of target atoms $k$.

The dominant error mechanisms can be grouped into four categories:
(i) dissipation-induced errors,
(ii) motional and Doppler-induced dephasing,
(iii) residual interactions between target atoms,
and (iv) technical noise.

\begin{figure*}[t]
   \centering
   \includegraphics[width=0.9\textwidth]{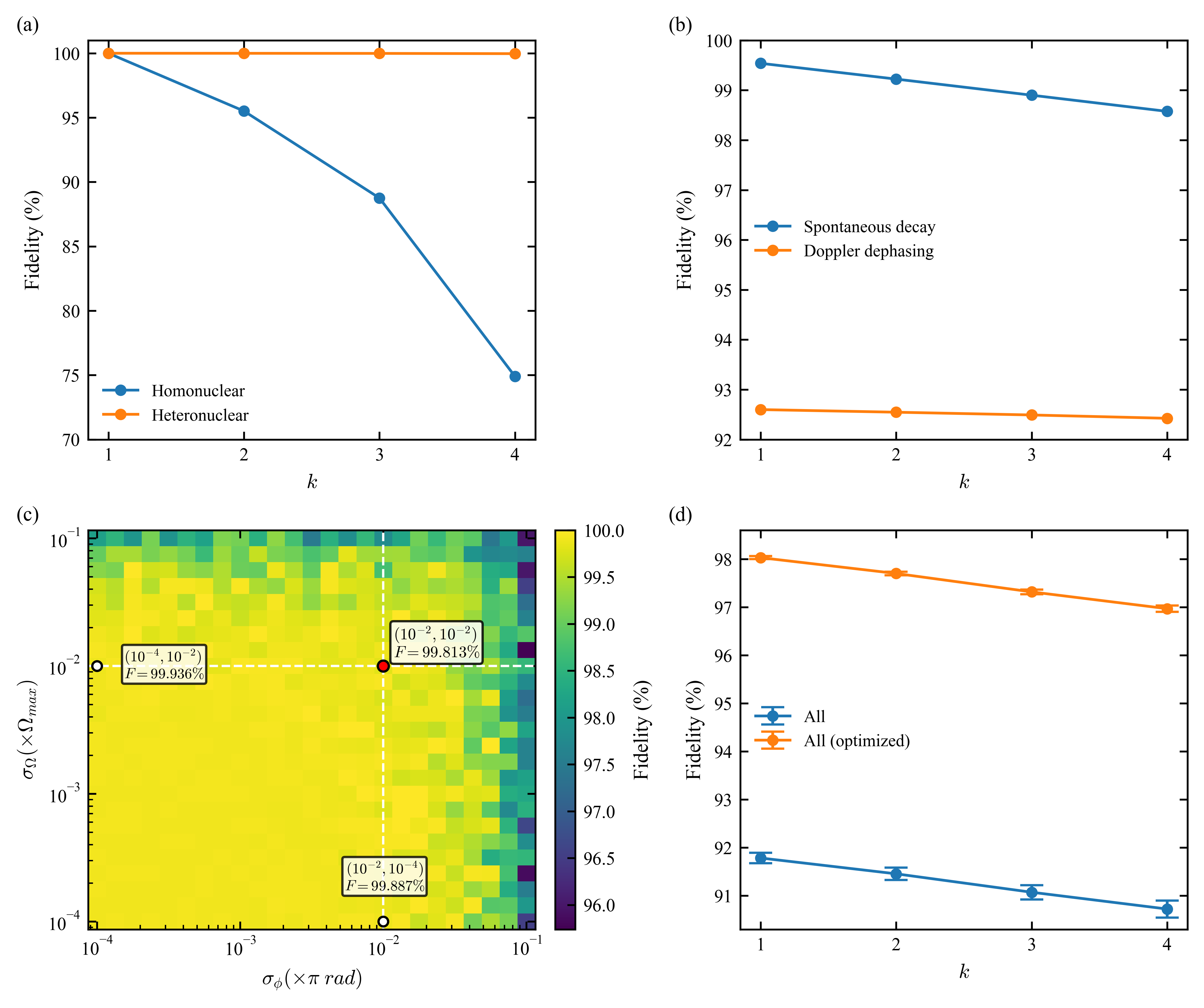}
   \caption{
   Summary of the dominant error sources and their scaling with the
   number of target qubits $k$. (a) Ideal gate fidelity versus $k$ for
   homonuclear and heteronuclear implementations, showing the strong
   suppression of errors induced by interactions between target atoms in the heteronuclear
   array. (b) Fidelity scaling under spontaneous decay and Doppler
   dephasing in the heteronuclear setting. (c) Fidelity map for the
   $k=1$ gate under combined technical amplitude and phase noise. The
   vertical and horizontal dashed lines indicate the technical-noise working
   point $\sigma_{\phi}=10^{-2}\pi~\mathrm{rad}$ and
   $\sigma_{\Omega}=10^{-2}\Omega_{\max}$ used later in the full-noise
   simulations. The marked points correspond to
   $(10^{-2},10^{-2})$ with $F=99.813\%$,
   $(10^{-2},10^{-4})$ with $F=99.887\%$, and
   $(10^{-4},10^{-2})$ with $F=99.936\%$, showing that amplitude and phase
   noise are nearly additive but exhibit a small positive cross contribution.
   (d) Gate fidelity under all noise channels, before and after optimization.}
   \label{fig:error_sources_summary}
\end{figure*}

Figure~\ref{fig:error_sources_summary} shows that errors induced by interactions
between target atoms, spontaneous emission, and Doppler dephasing all grow as
the number of target atoms increases, whereas the ideal
heteronuclear fidelity remains nearly independent of $k$ over the range
shown.  The technical-noise map in Fig.~\ref{fig:error_sources_summary}(c)
instead characterizes the single-target ($k=1$) gate as a function of the
specified laser-noise amplitudes.

\subsection{Dissipation-induced errors}
\label{subsec:dissipation}

Dissipative errors primarily arise from spontaneous emission of the
intermediate excited states.
Although the EIT/STIRAP protocol suppresses intermediate-state population in
the ideal coherent dark-state limit, dissipation, dephasing, and nonadiabatic
evolution perturb that limit and generate residual transient population.
Consequently, these open-system imperfections lead to irreversible scattering
and coherence loss.

For the $\mathrm{C}^1\mathrm{NOT}^k$ gate, the influence of spontaneous decay
increases with $k$ for two related reasons.
First, the total spontaneous-emission probability scales approximately with
the time-integrated population of the intermediate excited states, which
grows as more target atoms are driven simultaneously.
Second, residual interaction-induced perturbations of the target-atom dark state
enhance the admixture of excited states, thereby amplifying the effect of
spontaneous emission.
This trend is clearly visible in
Fig.~\ref{fig:error_sources_summary}(b), where the fidelity limited by
spontaneous decay decreases monotonically as $k$ increases.

\subsection{Motional and Doppler-induced dephasing}
\label{subsec:motional}

Finite atomic temperature gives rise to Doppler shifts in the excitation
lasers, which appear as random phase fluctuations in the ground--Rydberg
coherences.
In a multi-target configuration, the Doppler shifts experienced by different
target atoms are generally uncorrelated, so the accumulated random phases
grow with the size of the target register.

This effect is particularly detrimental for $\mathrm{C}^1\mathrm{NOT}^k$
gates, which rely on maintaining coherent dark-state evolution across all
targets simultaneously.
Even though the use of counterpropagating beams strongly suppresses the
first-order Doppler shift, the residual dephasing still imposes a
non-negligible fidelity ceiling.
As shown in Fig.~\ref{fig:error_sources_summary}(b), Doppler-induced
dephasing is already the dominant error channel in the heteronuclear setup,
while its dependence on $k$ remains weaker than that of spontaneous decay.

\subsection{Residual interactions between target atoms}
\label{subsec:target-interaction}

In addition to the intended Rydberg blockade between the control atom and
each target atom, residual dipole--dipole interactions among target atoms
are generally present.
These interactions generate state-dependent energy shifts that perturb the
ideal EIT resonance condition.

When the control atom is initialized in $|0\rangle_c$, the target atoms are
supposed to remain in the EIT dark state and avoid Rydberg excitation.
Residual interactions between target atoms introduce an effective interaction-induced
detuning that compromises the dark-state condition, weakens the adiabatic
elimination of the intermediate state, and results in unwanted population
transfer and extra phase accumulation.
This mechanism becomes increasingly severe as $k$ increases.

This many-body limitation is precisely the reason why the fidelity of the
homonuclear implementation drops rapidly with system size, whereas the
heteronuclear implementation remains nearly ideal, as shown in
Fig.~\ref{fig:error_sources_summary}(a).
The comparison demonstrates that separating the interaction scales of pairs
consisting of the control atom and a target atom from those of target-atom
pairs is essential for obtaining a scalable multi-target gate.

When the control atom is prepared in $|1\rangle_c$, the Rydberg blockade between
the control atom and each target atom strongly detunes the target-atom Rydberg level.
In this branch, the detrimental influence of interactions between target atoms is
partially masked by the stronger intended blockade, so their effect is less
severe than in the transparent branch.

\subsection{Technical noise}
\label{subsec:technical-noise}

In addition to intrinsic many-body effects, technical noise further limits
the achievable gate fidelity in realistic experiments.
Although these noise sources do not originate from the interacting
many-body structure itself, they do not scale favorably with system size and
therefore become increasingly relevant for multi-target gates.

\subsubsection{Laser intensity and detuning fluctuations}

Fluctuations of the excitation laser intensities lead to temporal variations
in the Rabi frequencies, thereby perturbing both the EIT condition and the
adiabatic evolution path.
As the number of target atoms increases, the sensitivity of the gate to
such fluctuations is enhanced, producing both population leakage and
coherent phase errors.

Fluctuations in the two-photon detuning may arise from magnetic-field noise
or laser-frequency drift.
They break the exact resonance condition required for dark-state protection,
leading to increased excited-state population and reduced fidelity.

\subsubsection{Laser phase noise}

Laser phase noise introduces stochastic phase fluctuations in the optical
driving fields.
In a multi-qubit setting, such fluctuations generate random relative phases
among different atoms and thereby degrade the global coherence required for
multi-target gate operation.

The combined effect of amplitude and phase noise is summarized in
Fig.~\ref{fig:error_sources_summary}(c), which shows the fidelity map in the
$(\sigma_{\phi},\sigma_{\Omega})$ plane, with $\sigma_{\phi}$ measured in
units of $\pi~\mathrm{rad}$ and $\sigma_{\Omega}$ in units of
$\Omega_{\max}$. Here $\sigma_{\phi}$ is the root-mean-square (r.m.s.)
phase fluctuation applied to the laser fields, and $\sigma_{\Omega}$ is the
r.m.s. fluctuation of the Rabi-frequency amplitude; both quantities are
defined relative to the units stated above.
The white dashed lines indicate the representative noise levels
$\sigma_{\phi}=10^{-2}$ and $\sigma_{\Omega}=10^{-2}$, whose intersection
(red marker) is the technical-noise working point adopted later in the
simulations that include all noise channels simultaneously.  The map and the
quoted value refer to the $k=1$ gate with amplitude and phase noise included
simultaneously (and with the other noise channels omitted): at this reference
point its fidelity is $99.813\%$.

The two additional marked points on the lower and left boundaries isolate
the effects of amplitude noise and phase noise separately:
$(10^{-2},10^{-4})$ gives $F=99.887\%$, while
$(10^{-4},10^{-2})$ gives $F=99.936\%$.
Expressed in terms of infidelity,
\begin{align}
\varepsilon_{\Omega} &= 1-0.99887 \approx 1.13\times 10^{-3},\\
\varepsilon_{\phi}   &= 1-0.99936 \approx 6.39\times 10^{-4},
\end{align}
whereas the combined-noise point corresponds to
\begin{align}
\varepsilon_{\phi,\Omega}=1-0.99813\approx 1.87\times 10^{-3}.
\end{align}
If the two technical-noise channels contributed independently and only at
leading order, one would expect
$\varepsilon_{\phi,\Omega}\approx\varepsilon_{\phi}+\varepsilon_{\Omega}$,
which gives $1.77\times 10^{-3}$, corresponding to a fidelity of about
$99.823\%$.
The actual value at the reference point is slightly lower, indicating a
small positive cross term,
\begin{equation}
\varepsilon_{\phi,\Omega}
=
\varepsilon_{\phi}
+
\varepsilon_{\Omega}
+
\varepsilon_{\mathrm{cross}},
\qquad
\varepsilon_{\mathrm{cross}}
\approx 1.0\times 10^{-4}.
\end{equation}
Therefore, the two technical-noise channels are approximately additive in
the weak-noise regime considered here, but not exactly so: a weak joint
effect is already visible once both fluctuations are present simultaneously.

When all noise channels are included simultaneously, the gate fidelity shows
a clear decrease with increasing $k$, as displayed by the blue curve in
Fig.~\ref{fig:error_sources_summary}(d).
However, the optimized protocol substantially suppresses this deterioration,
as shown by the orange curve, and preserves fidelities above $96\%$ even for
$\mathrm{C}^1\mathrm{NOT}^4$.
This comparison demonstrates that waveform optimization is essential once
all realistic noise sources are taken into account.

%%%%%%%%%%%%%%%%%%%%%%%%%%%%%%%%%%%%%%%%%%%%%%%%%%
\section{Error Suppression Strategies}
\label{sec:opt}

Based on the error analysis presented in Sec.~\ref{sec:dominant-errors},
we now discuss practical strategies for mitigating the dominant error
mechanisms in the $\mathrm{C}^1\mathrm{NOT}^k$ gate.
Rather than modifying the logical structure of the protocol,
these approaches aim to suppress error accumulation by engineering
interactions and control waveforms.
We focus on two complementary techniques:
(i) the use of heteronuclear Rydberg arrays to suppress residual
interactions between target atoms, and
(ii) optimal control of laser pulses to reduce non-adiabatic and
dissipative errors.

\subsection{Dual-species Rydberg atoms}
\label{subsec:dual_species}

As identified in Sec.~\ref{subsec:target-interaction},
a major limitation to the fidelity of the $\mathrm{C}^1\mathrm{NOT}^k$ gate
arises from residual interactions between target atoms, which perturb the
EIT dark-state condition and induce additional dissipation and dephasing.
An effective strategy to suppress this error mechanism is therefore to
engineer a strong separation between the interaction strengths associated
with the intended control-atom--target-atom interaction and the residual
interactions between target atoms.

A particularly natural way to realize such an interaction hierarchy is to
employ a heteronuclear Rydberg array, in which the control and target atoms
are encoded in different atomic species.
Due to their distinct atomic structures, heteronuclear and homonuclear atom
pairs can exhibit substantially different van der Waals coefficients.
As a result, the ratio
\begin{equation}
\eta \equiv \frac{|V_{ct}|}{|V_{tt}|}
\end{equation}
can be made much larger than in homonuclear configurations while keeping the
overall interaction scale within an experimentally accessible range.

From the viewpoint of the error mechanisms discussed in
Sec.~\ref{sec:dominant-errors}, increasing $\eta$ directly suppresses the
interaction-induced effective detuning experienced by the target atoms when
the control atom is in $|0\rangle_c$.
Consequently, the EIT dark-state condition is restored to a much higher
degree, leading to reduced transient excited-state population and, in turn,
less spontaneous scattering.
Importantly, this mitigation mechanism does not rely on delicate pulse-shape
fine-tuning, but instead exploits an intrinsic separation of interaction
energy scales.

The effectiveness of this approach is already evident in
Fig.~\ref{fig:error_sources_summary}(a), where the heteronuclear
implementation remains essentially ideal as $k$ increases, in sharp contrast
to the homonuclear case.
Specific atomic species, Rydberg levels, and geometric parameters that
realize large interaction ratios in practice are discussed in
Sec.~\ref{sec:experimental_mapping}.

\subsection{Optimal control}
\label{subsec:optimal_control}

Even with optimized interaction engineering, the fidelity of the
$\mathrm{C}^1\mathrm{NOT}^k$ gate remains limited by nonadiabatic errors,
dissipation, and technical noise during finite-time pulse sequences.
To further suppress these effects under realistic experimental constraints,
we employ optimal control to shape not only the Rabi frequencies but also the
effective detuning trajectories.

In our implementation, the optimization is performed around a physically
motivated reference pulse sequence rather than over completely unconstrained
waveforms.
Specifically, the control fields are decomposed into baseline pulses and
multiplicative correction functions,
\begin{align}
\Omega_\alpha(t)&=\Omega_{\alpha,0}(t)\,g_\alpha(t),\\
\delta_\beta(t)&=\delta_{\beta,0}(t)\,g_\beta(t),\\
\Delta_\beta(t)&=\Delta_{\beta,0}(t)\,g_\beta(t),
\end{align}
where $\alpha$ labels the Rabi-frequency channels and $\beta$ labels the
atomic detuning channels.
For the present protocol, these channels include the two STIRAP pulses of the
control atom, the probe and coupling pulses of the target atoms, and the
corresponding detuning modulations.
This parametrization preserves the overall temporal structure and boundary
conditions of the original pulse sequence, while still allowing enough freedom
to reduce leakage and improve robustness.

We use the standard dCRAB (dressed Chopped Random Basis) optimal-control
scheme~\cite{DCRAB1,DCRAB2}: each correction is represented by a small random
Fourier basis, and the basis is refreshed between optimization rounds to
escape restrictions of a fixed parametrization.  Explicitly, within a given
round we use
\begin{align}
g_\mu(t)
=
1+\lambda_\mu(t)
\sum_{n=1}^{N_c}
\left[
A_n^{(\mu)}\cos(\omega_n^{(\mu)} t)
+
B_n^{(\mu)}\sin(\omega_n^{(\mu)} t)
\right],
\label{eq:crab_expansion}
\end{align}
where $\mu$ labels the optimized channel,
$\lambda_\mu(t)$ is a smooth envelope vanishing at the beginning and end of the
corresponding pulse window, and the frequencies
$\omega_n^{(\mu)}=2\pi(n+r_n^{(\mu)})/T_\mu$
contain random offsets $r_n^{(\mu)}\in[0,1]$.
The envelope ensures that the optimized pulse remains anchored to the desired
initial and final conditions, while the randomized Fourier basis provides a
flexible but low-dimensional search space.

To maintain experimental feasibility, the optimized controls are not allowed to
grow arbitrarily.
After the multiplicative correction is applied, each waveform is passed through
a smooth saturation map that confines it to a prescribed interval,
\begin{equation}
u(t)\in [u_{\min},u_{\max}],
\end{equation}
with $u$ standing for either a Rabi frequency or a detuning.
Compared with hard clipping, this smooth saturation avoids introducing
nondifferentiable kinks into the waveform and leads to more stable numerical
optimization while still enforcing realistic amplitude bounds.
We additionally constrain the waveform slopes.  This bandwidth-motivated
condition excludes rapidly varying solutions that would be difficult to
realize with experimental pulse shaping hardware.

With this parametrization, the gate infidelity becomes a multivariate function
of the Fourier coefficients,
\begin{equation}
\mathcal{I}=1-F=\mathcal{I}\bigl(\{A_n^{(\mu)},B_n^{(\mu)}\}\bigr),
\end{equation}
which is minimized using a gradient-free Nelder--Mead algorithm.
A gradient-free search is particularly suitable here because the objective
function is obtained from repeated open-system time evolution under the
Lindblad master equation, making it computationally expensive and generally
ill-suited to stable gradient evaluation.
Depending on the simulation task, the target functional is chosen either as the
average gate infidelity or as the Bell-state infidelity of the entangled output
state.

The refreshed-basis dCRAB procedure retains successful corrections from
earlier rounds while enlarging the accessible control landscape.  This
standard construction is useful here because the open-system objective,
which includes dissipation, dephasing, and waveform-dependent leakage, is
rugged.

An important feature of the present implementation is that the effective
detunings are optimized on the same footing as the Rabi frequencies.
As discussed in Appendix~\ref{app:phase_detuning_equiv}, this is equivalent to
optimizing the time-dependent phases of the underlying complex laser couplings.
The control optimization can therefore be viewed as a unified reshaping of pulse
amplitudes, effective frequency chirps, and relative phases.

Figure~\ref{fig:optimized_results} makes the comparison protocol explicit.
The pre-optimized reference pulses in the insets of panels (a) and (d) are not
the original trial pulses.  They are obtained after optimizing the peak optical
resources and gate duration, and therefore provide a resource-optimized
baseline for the subsequent dCRAB waveform search.  Panels (a), (b), (d), and
(e) display the dCRAB Rabi frequencies and effective detunings, together with the
pre-optimized pulse envelopes used as the reference.

Panels (c) and (f) compare the all-noise dynamics of the $k=4$ gate.
The curves are averages over eight independent technical-noise realizations,
each evaluated with 128 quantum trajectories.  The dCRAB controls were obtained
for the $k=2$ gate and transferred without further optimization to the $k=4$
validation.  The resulting
phase-corrected Bell-state fidelity is $96.55(2.05)\%$ for dCRAB and
$95.30(3.55)\%$ for the pre-optimized reference, where parentheses give one
sample standard deviation.  dCRAB also lowers the transient population in the
intermediate manifolds and the integrated intermediate- and Rydberg-state
populations.  These reductions identify the waveform-induced suppression of
dissipative exposure, beyond the prior optimization of optical resources and
gate duration, as the mechanism responsible for the improved mean fidelity.

%%%%%%%%%%%%%%%%%%%%%%%%%%%%%%%
\section{Experimental mapping and feasibility}
\label{sec:experimental_mapping}

This section maps the proposed $\mathrm{C}^1\mathrm{NOT}^k$ protocol onto a
realistic heteronuclear neutral-atom platform.  We identify an experimentally
accessible interaction hierarchy, calibrate the dissipative parameters against
reported coherence measurements, and summarize the operating regime for a
two-species optical-tweezer implementation.

\subsection{Experimental platform and physical mapping}
\label{subsec:exp_platform}

As a representative platform, we consider a heteronuclear Rb--Cs
optical-tweezer array~\cite{2d-array1,experimental-setup,experimental-setup3,imperfection1,anand_dual-species_2024,petrosyan_fast_2024}.
The control qubit is encoded in a cesium atom and the target qubits in
rubidium atoms. This assignment allows the interaction between the control atom
and each target atom to be engineered independently of interactions between
target atoms through the choice of Rydberg states.

The dissipative parameters in the numerical simulations are calibrated to the
coherence data reported in Ref.~\cite{experimental-setup2}.  For the Rydberg
and intermediate excited states, we use the lifetimes
$T_1 = 88~\mu\mathrm{s},\tau_e = 110~\mathrm{ns}$
which correspond to the spontaneous-decay rates
\begin{align}
\Gamma_r = \frac{1}{T_1}
\approx 1.14\times10^4~\mathrm{s}^{-1}
\approx 2\pi\times 1.81~\mathrm{kHz}.\\
\Gamma_e=\frac{1}{\tau_e}
\approx 9.09\times10^6~\mathrm{s}^{-1}
\approx 2\pi\times1.45~\mathrm{MHz}.
\end{align}
In the open-system simulations, intermediate-state decay is the dominant
dissipative channel.  Direct Rydberg-state decay is much weaker over the gate
timescale.

The measured ground--Rydberg coherence time,
$T_2^* = 3~\mu\mathrm{s}$, defines the experimentally relevant dephasing scale
\begin{equation}
\frac{1}{T_2^*}
\approx 3.33\times10^5~\mathrm{s}^{-1}
\approx 2\pi\times53.1~\mathrm{kHz}.
\end{equation}
This dephasing includes finite-temperature effects, residual Doppler
broadening, laser light-shift fluctuations, and electric-field noise.
Following Ref.~\cite{experimental-setup2}, we represent these contributions by
an effective ground--Rydberg dephasing channel calibrated by the measured
$T_2^*$.  Because the numerical dephasing rate depends on the normalization
of the Lindblad jump operator, $T_2^*$ is used as the primary calibrated
quantity throughout the simulations.

\subsection{Interaction engineering and parameter selection}
\label{subsec:exp_interaction}

To suppress unwanted interactions between target atoms while retaining a strong
conditional blockade between the control atom and each target atom, we employ a
heteronuclear Rb--Cs Rydberg configuration and select the principal quantum
numbers by scanning the interaction coefficients with the Alkali Rydberg
Calculator (ARC)~\cite{beterov_rydberg_2015,ireland_interspecies_2024}.

In the parameter set considered here, the target atoms are Rb atoms excited to
$n_t=48$, whereas the control atom is a Cs atom excited to $n_c=51$.
For this choice, the van der Waals coefficients are
\begin{equation}
C_6^{ct}=-763.8~\mathrm{GHz}\cdot\mu\mathrm{m}^6,
\\
C_6^{tt}=-9.306~\mathrm{GHz}\cdot\mu\mathrm{m}^6,
\end{equation}
so that the interaction ratio is
\begin{equation}
\eta=\left|\frac{C_6^{ct}}{C_6^{tt}}\right|
\approx 82.
\end{equation}
This large ratio is essential for the multi-target EIT gate. The interaction
between the control atom and a target atom breaks the EIT condition when the
control atom occupies its Rydberg state, whereas residual interactions between target atoms produce
only a small perturbation of the desired dark-state evolution.

Because $V(R)=C_6/R^6$, the absolute interaction scale is tunable through the
tweezer spacing.  After choosing atomic species and Rydberg levels that yield
a favorable $|C_6^{ct}/C_6^{tt}|$, the interatomic distance can therefore be
used to select a practical working point with strong conditional blockade and
weak residual interactions between target atoms.

Atoms in realistic tweezer arrays fluctuate around their equilibrium positions
by approximately $0.1~\mu\mathrm{m}$. The large contrast between the
control-atom--target-atom and target-atom--target-atom $C_6$ coefficients means that moderate
position fluctuations do not qualitatively alter the required interaction
hierarchy.  The heteronuclear Rb--Cs implementation should therefore remain
feasible in the presence of realistic motional uncertainty.

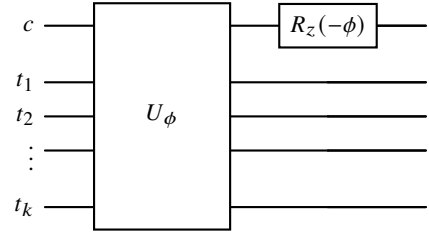
\begin{figure}[!t]
\centering
\begin{quantikz}[row sep=0.45cm, column sep=0.65cm]
\lstick{$c$}   & \gate[5][1.8cm]{U_{\phi}} & \gate{R_z(-\phi)} & \qw \\
\lstick{$t_1$} &                             &                    & \qw \\
\lstick{$t_2$} &                             &                    & \qw \\
\lstick{$\vdots$} &                          &                    & \qw \\
\lstick{$t_k$} &                             &                    & \qw
\end{quantikz}
\caption{Circuit-level representation of the implemented multi-target gate. The
native operation is $U_{\phi}$ defined in Eq.~\eqref{eq:Uphi}. Since the extra
phase $\phi$ is a local phase on the control qubit, it can be compensated by a
single-qubit $R_z(-\phi)$ rotation, yielding the standard
$\mathrm{C}^1\mathrm{NOT}^k$ gate.}
\label{fig:circuit}
\end{figure}

\begin{figure*}[t]
   \centering
   \includegraphics[width=0.96\textwidth]{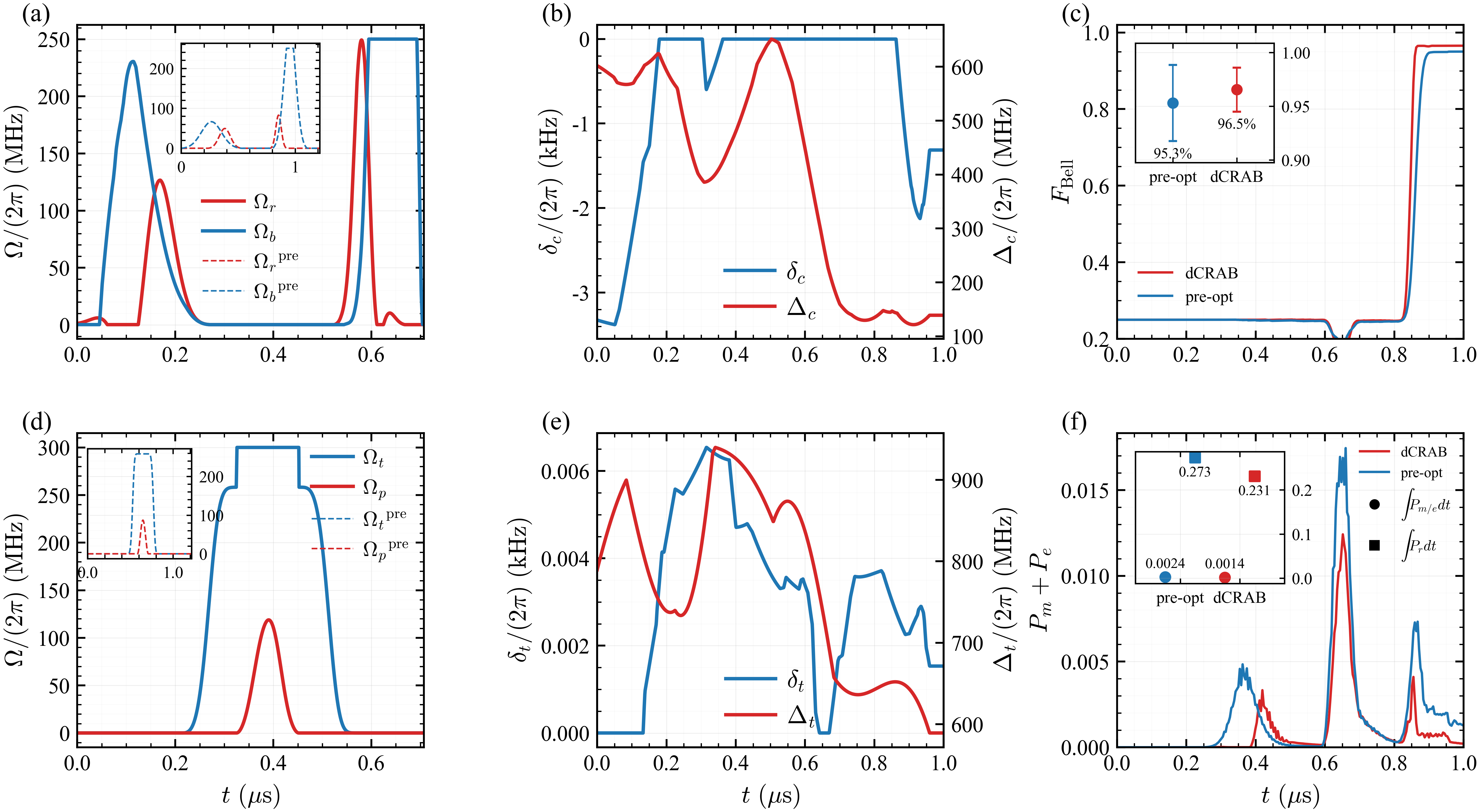}
   \caption{
   dCRAB waveform control and all-noise dynamical validation for the
   heteronuclear $\mathrm{C}^1\mathrm{NOT}^{4}$ operation. (a) dCRAB
   control-atom Rabi frequencies $\Omega_r$ and $\Omega_b$. (b) Corresponding
   control-atom effective detunings $\delta_c$ and $\Delta_c$. (c)
   Phase-corrected Bell-state fidelity during the all-noise $k=4$ operation.
   (d) dCRAB target-atom Rabi frequencies $\Omega_t$ and $\Omega_p$.
   (e) Corresponding target-atom effective detunings $\delta_t$ and
   $\Delta_t$. Insets in (a) and (d) show the pre-optimized reference
   envelopes. This reference already includes optimization of the peak optical
   resources and gate duration. (f) Corresponding total
   intermediate-manifold population, $P_m+P_e$. In (c) and (f), blue and red
   curves denote the pre-optimized and dCRAB controls, respectively. The inset
   in (c) reports the final fidelity; the inset in (f) compares the integrated
   intermediate- and Rydberg-state populations, denoted by circles and squares,
   respectively. The dCRAB control fields were optimized for
   $k=2$ and transferred without further optimization to $k=4$.
   Curves are averages over eight independent technical-noise realizations at
   $\sigma_{\phi}=10^{-2}\pi~\mathrm{rad}$ and
   $\sigma_{\Omega}=10^{-2}\Omega_{\max}$, with 128 quantum trajectories per
   realization. Error bars in the inset of (c) indicate one sample standard
   deviation.}
   \label{fig:optimized_results}
\end{figure*}

\subsection{Circuit-level interpretation}
\label{subsec:circuit}

As illustrated in Fig.~\ref{fig:circuit}, the protocol implements the following
multi-target controlled operation:
\begin{equation}
U_{\phi}
=
|0\rangle\!\langle 0|_c \otimes I
+
\mathrm{e}^{i\phi}
|1\rangle\!\langle 1|_c \otimes X^{\otimes k},
\label{eq:Uphi}
\end{equation}
where $k$ is the number of target qubits and $\phi$ is an additional phase
accumulated during the gate.

This is a local phase on the control qubit and can therefore be removed by a
single-qubit $R_z$ rotation.  Equation~\eqref{eq:Uphi} can be rewritten as
\begin{equation}
U_{\phi}
=
\left(
|0\rangle\!\langle 0|
+
\mathrm{e}^{i\phi}|1\rangle\!\langle 1|
\right)_c
\,U,
\label{eq:Uphi_decomp}
\end{equation}
where $U \equiv U_{\mathrm{C}^1\mathrm{NOT}^k}$ denotes the standard multi-target
controlled-$X$ gate. Hence, by applying the inverse phase gate
$\mathrm{diag}(1,\mathrm{e}^{-i\phi})$ to the control qubit, one recovers the
standard $\mathrm{C}^1\mathrm{NOT}^k$ operation.

Up to a physically irrelevant global phase, the inverse phase gate is equivalent to
a single-qubit rotation $R_z(-\phi)$, since
\begin{equation}
R_z(-\phi)
=
\begin{pmatrix}
\mathrm{e}^{i\phi/2} & 0\\
0 & \mathrm{e}^{-i\phi/2}
\end{pmatrix},
\end{equation}
which differs from $\mathrm{diag}(1,\mathrm{e}^{-i\phi})$ only by the global
phase factor $\mathrm{e}^{i\phi/2}$.  Thus, an $R_z(-\phi)$ gate on the
control qubit compensates the additional phase.

We also treat $\phi$ as an optimization variable.  This enlarges the accessible
control landscape and can yield simpler, more experimentally feasible
waveforms.  With otherwise identical optimized pulses, changing $k$ can induce
an additional $\pi$ shift in $\phi$.  The Bell-type state obtained from an
initial $|+\rangle_c$ control qubit consequently changes from
$(|00\rangle+|11\rangle)/\sqrt{2}$ to
$(|00\rangle-|11\rangle)/\sqrt{2}$.  The dynamical origin of this
$k$-dependent phase shift is analyzed in
Appendix~\ref{app:parity_phase}.

\subsection{Optimized performance and feasibility}
\label{subsec:exp_summary}

Heteronuclear interaction engineering and waveform optimization jointly
suppress interaction-induced, dissipative, and technical errors.  Before the
waveform search, we perform a pre-optimization using the gradient-free
Nelder--Mead direct-search method.  This low-dimensional stage jointly varies
the gate duration, peak Rabi-frequency scales, detunings, interatomic spacing,
and the additional control-qubit phase $\phi$ to minimize the chosen gate
infidelity.  It therefore establishes an experimentally informed reference,
rather than retaining the original trial pulses.

Starting from this pre-optimized reference, dCRAB introduces constrained
Fourier corrections to the Rabi frequencies and effective detunings.  The
amplitude and slope constraints prevent unphysical peak powers and highly
irregular waveforms.  As shown in Fig.~\ref{fig:optimized_results}(a), (b),
(d), and (e), the peak waveform strengths and pulse durations remain close to
their pre-optimized values, while the temporal shapes and effective detuning
trajectories acquire additional structure.  The fidelity improvement therefore
arises from the additional waveform degrees of freedom, rather than from a
wholesale change in the operating time or optical intensity scale.

The $k=2$ dCRAB controls are transferred without further optimization to the
$k=4$ gate.  At the all-noise working point, the phase-corrected Bell-state
fidelity in Fig.~\ref{fig:optimized_results}(c) is
$96.55(2.05)\%$ for dCRAB and $95.30(3.55)\%$ for the pre-optimized reference.
These values are means over eight independent technical-noise realizations,
each evaluated with 128 quantum trajectories, and parentheses denote one
sample standard deviation.  The higher mean and smaller standard deviation for
dCRAB demonstrate enhanced robustness to stochastic technical noise, even
after transfer to the larger $k=4$ target register.

Panel (f) resolves the associated population dynamics.  The dCRAB waveform
reduces the transient population in the intermediate manifolds and lowers the
integrated intermediate-state population from $2.4\times10^{-3}$ to
$1.4\times10^{-3}~\mu\mathrm{s}$.  It also lowers the integrated
Rydberg-state population from $0.273$ to $0.231~\mu\mathrm{s}$.  The reduced
population exposure is consistent with less spontaneous scattering and
Rydberg dephasing.  The improved mean fidelity therefore accompanies temporal
waveform shaping beyond the pre-optimization of gate duration and optical
resources.

These results indicate that the proposed $\mathrm{C}^1\mathrm{NOT}^k$ protocol
is compatible with experimentally realistic neutral-atom parameters.  It thus
provides a viable route to native multi-qubit gates for quantum error
correction.  The experimental records used to calibrate the optical bounds
and noise scales are compiled in
Appendix~\ref{app:experimental_parameter_sets}.

%%%%%%%%%%%%%%%%%%%%%%%%%%%%%%%%%%%%
\section{Conclusions} \label{sec:conclusion}

We investigated an EIT-based route to native $\mathrm{C}^1\mathrm{NOT}^k$
gates in neutral-atom Rydberg arrays, focusing on the four-target operation
relevant to weight-4 stabilizer readout.  A native
$\mathrm{C}^1\mathrm{NOT}^k$ gate can compress the entangling layer of a
weight-$k$ stabilizer measurement from $\mathcal{O}(k)$ sequential two-qubit
operations to a constant-depth block.

We formulated the microscopic Hamiltonian and open-system dynamics of the
protocol, and identified its dominant error mechanisms.  The principal
limitations are spontaneous emission from intermediate states,
Doppler-induced dephasing, and residual interactions between target atoms.  The
last mechanism perturbs the EIT dark state and accounts for the rapid fidelity
loss of the homonuclear implementation as the number of target atoms grows.

To address this bottleneck, we combined heteronuclear interaction engineering
with waveform optimization. The Rb--Cs configuration separates the interaction
scales of control-atom--target-atom pairs and target-atom--target-atom pairs, thereby restoring
dark-state protection in the multi-target setting.  dCRAB-based optimal
control then optimizes the Rabi-frequency profiles and effective detunings,
reducing nonadiabatic leakage, spontaneous scattering, and sensitivity to
technical noise.  The resource-matched comparison with the pre-optimized
reference further shows that the additional gain is obtained through waveform
reshaping and reduced intermediate- and Rydberg-state exposure, beyond the
prior optimization of peak optical resources and gate duration.

With experimentally realistic parameters and all major noise channels
included, the optimized heteronuclear protocol reaches fidelities of
$98.03\%$, $97.70\%$, $97.32\%$, and $96.54\%$ for
$\mathrm{C}^1\mathrm{NOT}^{1}$ through $\mathrm{C}^1\mathrm{NOT}^{4}$,
respectively.  The residual gate phase is local to the control qubit and can
be removed by a single-qubit $R_z$ correction.  Its parity-dependent $\pi$
shift arises from the effective $-X$ operation accumulated by each target
atom in the blockade branch.  Finally, for the control-atom two-photon
process, optimizing the effective detunings is equivalent to optimizing the
phases of the complex Rabi frequencies.

Taken together, these results support EIT-based native multi-target gates as a
promising primitive for neutral-atom quantum information processing.  Beyond
the Toric Code, the framework applies to a broader class of stabilizer
measurements and multi-qubit entangling operations.  Future work should test
larger target registers, explicitly fault-tolerant stabilizer-extraction
circuits, and the effects of correlated noise and atom loss.

\begin{acknowledgments}
We thank Yibo Wang, Kunpeng Wang, Leiyinan Liu, and Zhifeng
Zhao for valuable discussions and technical guidance. This work is supported by the National Key Research and Development
of China (Grant Nos. 2021YFA1402001), the National Natural Science Foundation
of China (NSFC) (Grant Nos. 12375007), the Strategic Priority Research Program
of the Chinese Academy of Sciences (Grant No. XDB1690000), and the National
Innovation Program for Quantum Science and Technology of China (Grant No.
2023ZD0300401). 
\end{acknowledgments}

% 指定以下部分为附录。如果只有一个附录，请使用 \appendix*。
\appendix

\section{Detailed dynamics of EIT dark-state evolution}
\label{app:eit_dynamics}

This appendix derives the EIT dark-state dynamics governing the target-atom
evolution in the gate protocol, following the mechanism proposed by
M\"uller \emph{et al.}~\cite{EIT-protocol}.

%\subsection{Target Hamiltonian and symmetry reduction}
%\label{app:eit_symmetry}

We start from the target Hamiltonian in the rotating frame [main text
Eq.~\eqref{eq:Ht}]:
\begin{align}
\mathcal{H}_{t}(t)
&=
\frac{\hbar\Omega_p(t)}{2}
\Bigl(|0\rangle\langle e|+|1\rangle\langle e|\Bigr)
+\frac{\hbar\Omega_c(t)}{2}\,|e\rangle\langle r|
+\mathrm{h.c.}
\nonumber\\
&\qquad
-\hbar\Delta_t\,|e\rangle\langle e|
-\hbar\delta_t\,|r\rangle\langle r|.
\label{eq:Ht_app}
\end{align}

Introduce the symmetric and antisymmetric qubit states
\begin{align}
|d_1\rangle
=
|-\rangle
=
\frac{|1\rangle-|0\rangle}{\sqrt{2}},
\qquad
|+\rangle
=
\frac{|0\rangle+|1\rangle}{\sqrt{2}} .
\label{eq:plus_d1_app}
\end{align}
Using
\begin{align}
|0\rangle\langle e|+|1\rangle\langle e|
=
\sqrt{2}\,|+\rangle\langle e|,
\end{align}
we see that the probe field couples only $|+\rangle$ to $|e\rangle$, while
$|d_1\rangle$ is decoupled:
\begin{align}
\mathcal{H}_{t}(t)\,|d_1\rangle=0
\qquad
\text{(within the driven manifold)}.
\end{align}
The dynamics therefore separates into a trivial one-dimensional sector,
$\{|d_1\rangle\}$, and a three-level $\Lambda$ sector spanned by
$\{|+\rangle,|e\rangle,|r\rangle\}$.

In the ordered basis $\{|+\rangle,|e\rangle,|r\rangle\}$, the Hamiltonian reads
\begin{align}
\mathcal{H}_{\Lambda}(t)
=
\hbar
\begin{pmatrix}
0 & \Omega_p(t)/\sqrt{2} & 0\\
\Omega_p(t)/\sqrt{2} & -\Delta_t & \Omega_c(t)/2\\
0 & \Omega_c(t)/2 & -\delta_t
\end{pmatrix},
\label{eq:H_lambda_matrix}
\end{align}
where a real gauge has been chosen such that $\Omega_p(t)$ and $\Omega_c(t)$ are
real.

%\subsection{Instantaneous eigenstates}
%\label{app:eit_eigenstates}

We first consider two-photon resonance, $\delta_t=0$, which is the EIT
condition.  We seek an eigenstate with a vanishing excited-state component,
\begin{align}
|d_2(t)\rangle
=
a(t)\,|+\rangle+b(t)\,|r\rangle,
\qquad
\langle e|d_2(t)\rangle=0 .
\end{align}
Imposing
$\mathcal{H}_{\Lambda}(t)|d_2(t)\rangle=E_{d_2}(t)|d_2(t)\rangle$
and using the $|e\rangle$ row gives
\begin{align}
\frac{b(t)}{a(t)}
=
-\frac{\sqrt{2}\,\Omega_p(t)}{\Omega_c(t)}
\equiv -x(t).
\end{align}
Defining the mixing angle $\theta(t)$ by
\begin{align}
\tan\theta(t)=x(t),
\end{align}
the normalized dark state becomes
\begin{align}
|d_2(t)\rangle
=
\cos\theta(t)\,|+\rangle-\sin\theta(t)\,|r\rangle.
\label{eq:dark_theta_app}
\end{align}
Substituting Eq.~\eqref{eq:dark_theta_app} into
Eq.~\eqref{eq:H_lambda_matrix} with $\delta_t=0$ yields
\begin{align}
\mathcal{H}_{\Lambda}(t)\,|d_2(t)\rangle=0,
\qquad
(\delta_t=0),
\label{eq:dark_eigenvalue_zero_app}
\end{align}
Thus, the dark eigenvalue vanishes in the rotating frame and the excited state
$|e\rangle$ remains unpopulated in the ideal adiabatic limit.

An orthonormal complement in $\mathrm{span}\{|+\rangle,|r\rangle\}$ is the bright
state
\begin{align}
|b(t)\rangle
=
\sin\theta(t)\,|+\rangle+\cos\theta(t)\,|r\rangle,
\qquad
\langle b(t)|d_2(t)\rangle=0.
\label{eq:bright_app}
\end{align}
In the adiabatic basis
$\{|d_2(t)\rangle,|b(t)\rangle,|e\rangle\}$, the dark state is decoupled from
$|e\rangle$ at the Hamiltonian level, while the bright manifold
$\{|b(t)\rangle,|e\rangle\}$ is coupled with an effective Rabi frequency
\begin{align}
\Omega_{\mathrm{eff}}(t)
=
\frac{1}{2}\sqrt{\Omega_c^2(t)+2\Omega_p^2(t)},
\label{eq:Omega_eff_app}
\end{align}
which sets the characteristic gap separating the dark and bright sectors (up to
detuning-dependent corrections through $\Delta_t$).

%\subsection{Adiabatic dynamics}
%\label{app:eit_adiabatic}

Assume that the EIT pulse is smooth and satisfies
$\Omega_p(0)=\Omega_p(T)=0$. Then $\theta(0)=\theta(T)=0$, so that
\begin{align}
|d_2(0)\rangle=|+\rangle,
\qquad
|d_2(T)\rangle=|+\rangle.
\end{align}
Under adiabatic evolution,
\begin{align}
|\psi(t)\rangle
\approx
e^{i\gamma(t)}|d_2(t)\rangle,
\label{eq:adiabatic_follow_app}
\end{align}
where the total phase $\gamma(t)$ contains both dynamical and geometric
contributions.

Because $E_{d_2}(t)=0$ at $\delta_t=0$
[Eq.~\eqref{eq:dark_eigenvalue_zero_app}], the dynamical phase vanishes:
\begin{align}
\gamma_{\mathrm{dyn}}(T)
=
-\frac{1}{\hbar}\int_0^T E_{d_2}(t)\,dt
=
0.
\end{align}
The Berry phase is
\begin{align}
\gamma_{\mathrm{geo}}(T)
=
i\int_0^T
\langle d_2(t)|\partial_t d_2(t)\rangle\,dt.
\end{align}
In the real gauge chosen above, one has
$\langle d_2(t)|\partial_t d_2(t)\rangle=0$, hence
\begin{align}
\gamma_{\mathrm{geo}}(T)=0.
\end{align}
Thus, in the EIT-intact branch, the $|+\rangle$ component returns with
negligible relative phase, while $|d_1\rangle$ remains unchanged.  This is a
transparent evolution of the target atom.

In the time-dependent adiabatic basis, a nonadiabatic coupling arises from the
time dependence of $\theta(t)$. Using
Eqs.~\eqref{eq:dark_theta_app} and \eqref{eq:bright_app}, one obtains
\begin{align}
\langle b(t)|\partial_t d_2(t)\rangle
=
-\dot{\theta}(t).
\end{align}
A sufficient adiabaticity condition is
\begin{align}
|\dot{\theta}(t)|\ll \Omega_{\mathrm{eff}}(t).
\label{eq:adiabaticity_app}
\end{align}
Since
\begin{align}
\tan\theta(t)
=
\frac{\sqrt{2}\Omega_p(t)}{\Omega_c(t)},
\end{align}
we have
\begin{align}
\dot{\theta}(t)
=
\frac{
\sqrt{2}\bigl[\Omega_c(t)\dot{\Omega}_p(t)-\Omega_p(t)\dot{\Omega}_c(t)\bigr]
}{
\Omega_c^2(t)+2\Omega_p^2(t)
}.
\label{eq:theta_dot_app}
\end{align}
In addition, a large one-photon detuning
$|\Delta_t|\gg |\Omega_p|,|\Omega_c|$
suppresses the intermediate-state population and thereby reduces spontaneous
scattering.

%\subsection{Blockade-induced EIT breaking}
%\label{app:eit_breaking_detailed}

During the EIT stage, the interaction between the control atom and a target atom shifts the target-atom
Rydberg level and modifies the effective two-photon detuning. Denoting the
control-atom Rydberg occupation by $n_{rc}\in\{0,1\}$, we write
\begin{align}
\delta_t^{(\mathrm{eff})}
=
\delta_t-\frac{V_{ct}(R)}{\hbar}\,n_{rc}.
\label{eq:delta_eff_app}
\end{align}
This leads to two qualitatively distinct dynamical branches.

When the control atom remains in $|0\rangle_c$, one has $n_{rc}=0$ and takes
$\delta_t^{(\mathrm{eff})}\approx0$. Each target atom then adiabatically follows the dark
state and returns to its qubit manifold with negligible excitation of
$|e\rangle_t$. This implements approximately the identity on the target atom during the
EIT stage.

When the control atom is excited to $|r\rangle_c$, the induced shift can be made large:
\begin{align}
\left|\delta_t^{(\mathrm{eff})}\right|
\sim
\frac{|V_{ct}(R)|}{\hbar}
\gg
\Omega_c(t),
\label{eq:blockade_condition_app}
\end{align}
so that $|r\rangle_t$ is far off resonance and the EIT pathway is effectively
closed. In this limit, the dynamics is dominated by the far-detuned coupling
between $\{|0\rangle_t,|1\rangle_t\}$ and $|e\rangle_t$ through $\Omega_p(t)$.

Adiabatically eliminating $|e\rangle_t$ to second order in $\Omega_p/\Delta_t$
gives an effective Hamiltonian in the ground-state manifold
$\{|0\rangle,|1\rangle\}$:
\begin{align}
\mathcal{H}_{\mathrm{eff}}^{(g)}(t)
\simeq
-\frac{\hbar\Omega_p^2(t)}{4\Delta_t}
\Bigl(
|0\rangle\langle 0|
+
|1\rangle\langle 1|
+
|0\rangle\langle 1|
+
|1\rangle\langle 0|
\Bigr).
\label{eq:Heff_ground_app}
\end{align}
Transforming to the $\{|+\rangle,|d_1\rangle\}$ basis yields
\begin{align}
\mathcal{H}_{\mathrm{eff}}^{(g)}(t)
\simeq
-\frac{\hbar\Omega_p^2(t)}{2\Delta_t}|+\rangle\langle +|
-\frac{\hbar\Omega_p^2(t)}{4\Delta_t}I_g,
\label{eq:Heff_plus_app}
\end{align}
where $I_g$ is the identity on the ground manifold. The second term contributes
only a global phase, whereas the first produces a relative phase on $|+\rangle$:
\begin{align}
|+\rangle \longrightarrow e^{i\phi}|+\rangle,
\qquad
\phi
=
\int_0^{T_p}\frac{\Omega_p^2(t)}{2\Delta_t}\,dt.
\label{eq:phi_app}
\end{align}
Choosing the pulse area such that $\phi=\pi$, equivalently
\begin{align}
\int_0^{T_p}\frac{\Omega_p^2(t)}{\Delta_t}\,dt
=
2\pi,
\label{eq:area_condition_app}
\end{align}
gives
\begin{align}
|+\rangle \longrightarrow -|+\rangle,
\qquad
|d_1\rangle \longrightarrow |d_1\rangle.
\label{eq:plus_flip_app}
\end{align}
Returning to the computational basis,
\begin{align}
|0\rangle
=
\frac{|+\rangle-|d_1\rangle}{\sqrt{2}},
\qquad
|1\rangle
=
\frac{|+\rangle+|d_1\rangle}{\sqrt{2}},
\end{align}
we obtain
\begin{align}
|0\rangle \longrightarrow -|1\rangle,
\qquad
|1\rangle \longrightarrow -|0\rangle,
\label{eq:X_from_phase_app}
\end{align}
which is a bit flip up to an overall phase.  This result shows how EIT breaking
converts the target-atom evolution into an effective $-X$ operation.

\section{Parity-dependent phase of the multi-target gate}
\label{app:parity_phase}

This appendix explains why the phase of the implemented
$\mathrm{C}^1\mathrm{NOT}^k$ gate acquires an additional $\pi$ when the parity
of $k$ changes under otherwise identical optimized pulses.

The key observation already follows from the single-target result
Eq.~\eqref{eq:X_from_phase_app}. In the blockade branch, when the control atom is
excited to $|r\rangle_c$ and the EIT condition is broken, the effective operation
on a \emph{single} target atom is
\begin{align}
|0\rangle \longrightarrow -|1\rangle,
\qquad
|1\rangle \longrightarrow -|0\rangle,
\end{align}
that is,
\begin{align}
U_t^{(1)}=-X.
\label{eq:minusX_single}
\end{align}

For $k$ target atoms driven simultaneously by the same control atom, and
neglecting for the moment corrections from interactions between target atoms, the target-atom
operation in the blockade branch factorizes as
\begin{align}
U_t^{(k)}
=
\bigl(U_t^{(1)}\bigr)^{\otimes k}
=
(-X)^{\otimes k}
=
(-1)^k X^{\otimes k}.
\label{eq:Uk_targets}
\end{align}
Hence the sign accumulated in the blockade branch depends only on the parity of
$k$.

Including the transparent branch, the full controlled operation may therefore be
written as
\begin{align}
U_k
=
|0\rangle\!\langle 0|_c\otimes I
+
e^{i\phi_0}(-1)^k
|1\rangle\!\langle 1|_c\otimes X^{\otimes k},
\label{eq:Uk_parity}
\end{align}
where $\phi_0$ collects all $k$-independent contributions, such as the common
phases accumulated in the control pulses and other shared dynamical phases. This
can be equivalently expressed as
\begin{align}
U_k
=
|0\rangle\!\langle 0|_c\otimes I
+
e^{i\phi_k}
|1\rangle\!\langle 1|_c\otimes X^{\otimes k},
\end{align}
with
\begin{align}
\phi_k
=
\phi_0+k\pi
\qquad
(\mathrm{mod}\ 2\pi).
\label{eq:phi_k_parity}
\end{align}
Therefore,
\begin{align}
\phi_{k+1}-\phi_k
=
\pi
\qquad
(\mathrm{mod}\ 2\pi),
\end{align}
showing that odd and even values of $k$ differ by an additional $\pi$ phase.

This parity effect is directly visible when the gate acts on an initial state
with the control atom prepared in $|+\rangle_c$ and the target atoms in
$|0\cdots 0\rangle_t$:
\begin{align}
|+\rangle_c\otimes |0\cdots0\rangle_t
\longrightarrow
\frac{
|0\rangle_c|0\cdots0\rangle_t
+
e^{i\phi_k}|1\rangle_c|1\cdots1\rangle_t
}{\sqrt{2}}.
\label{eq:bell_like_parity}
\end{align}
Because $\phi_k$ changes by $\pi$ when the parity of $k$ changes, the relative
sign of the Bell-type or GHZ-type superposition flips accordingly:
\begin{align}
\frac{|0\cdots0\rangle+|1\cdots1\rangle}{\sqrt{2}}
\longleftrightarrow
\frac{|0\cdots0\rangle-|1\cdots1\rangle}{\sqrt{2}}.
\end{align}

Equation~\eqref{eq:phi_k_parity} is the ideal parity argument. In the full
optimized dynamics, residual interactions between target atoms and other imperfections may
further shift $\phi_k$ by a common correction. Nevertheless, the additional $\pi$
phase associated with the parity of $k$ remains rooted in the fact that each target
contributes one factor of $-1$ through the blockade-induced effective $-X$
operation.

\section{Equivalence between detuning optimization and phase optimization in the control-qubit two-photon process}
\label{app:phase_detuning_equiv}

This appendix shows that, for the control-atom two-photon excitation process,
optimizing the one- and two-photon detunings is equivalent to optimizing the
phases of the complex Rabi frequencies.  This gauge equivalence transfers
time-dependent driving-field phases into diagonal detunings through a
time-dependent basis transformation, and vice versa.

%\subsection{Control-qubit Hamiltonian with complex Rabi frequencies}
%\label{app:control_phase_hamiltonian}

Consider the ladder-type three-level subspace
$\{|1\rangle_c,|m\rangle_c,|r\rangle_c\}$ of the control atom.  Let the two
Rabi frequencies be complex and time dependent:
\begin{align}
\Omega_r(t)
=
|\Omega_r(t)|\,e^{i\varphi_r(t)},
\qquad
\Omega_b(t)
=
|\Omega_b(t)|\,e^{i\varphi_b(t)}.
\end{align}
Then the rotating-frame Hamiltonian is
\begin{align}
\mathcal{H}_c(t)
&=
\frac{\hbar}{2}\Omega_r(t)\,|1\rangle_c\langle m|
+
\frac{\hbar}{2}\Omega_b(t)\,|m\rangle_c\langle r|
+
\mathrm{h.c.}
\nonumber\\
&\qquad
-\hbar\Delta_c(t)\,|m\rangle_c\langle m|
-\hbar\delta_c(t)\,|r\rangle_c\langle r|.
\label{eq:Hc_complex_phase}
\end{align}
In the ordered basis
$\{|1\rangle_c,|m\rangle_c,|r\rangle_c\}$, this becomes
\begin{align}
\mathcal{H}_c(t)
=
\hbar
\begin{pmatrix}
0 & |\Omega_r|e^{i\varphi_r}/2 & 0\\
|\Omega_r|e^{-i\varphi_r}/2 & -\Delta_c & |\Omega_b|e^{i\varphi_b}/2\\
0 & |\Omega_b|e^{-i\varphi_b}/2 & -\delta_c
\end{pmatrix}.
\label{eq:Hc_matrix_complex_phase}
\end{align}

%\subsection{Gauge transformation to real couplings}
%\label{app:control_phase_gauge}

Define the time-dependent diagonal unitary transformation
\begin{align}
W_c(t)
=
|1\rangle_c\langle 1|
+
e^{-i\varphi_r(t)}|m\rangle_c\langle m|
+
e^{-i[\varphi_r(t)+\varphi_b(t)]}|r\rangle_c\langle r|.
\label{eq:Wc_phase_gauge}
\end{align}
For the transformed state
$|\tilde{\psi}(t)\rangle=W_c^\dagger(t)|\psi(t)\rangle$, the Hamiltonian is
\begin{align}
\tilde{\mathcal{H}}_c(t)
=
W_c^\dagger \mathcal{H}_c W_c
-i\hbar W_c^\dagger \dot W_c .
\label{eq:Hc_tilde_def}
\end{align}
A direct calculation gives
\begin{align}
\tilde{\mathcal{H}}_c(t)
&=
\frac{\hbar}{2}|\Omega_r(t)|\,|1\rangle_c\langle m|
+
\frac{\hbar}{2}|\Omega_b(t)|\,|m\rangle_c\langle r|
+
\mathrm{h.c.}
\nonumber\\
&\qquad
-\hbar\tilde{\Delta}_c(t)\,|m\rangle_c\langle m|
-\hbar\tilde{\delta}_c(t)\,|r\rangle_c\langle r|,
\label{eq:Hc_real_phase}
\end{align}
with the effective detunings
\begin{align}
\tilde{\Delta}_c(t)
&=
\Delta_c(t)+\dot{\varphi}_r(t),
\label{eq:Deltac_shift}\\
\tilde{\delta}_c(t)
&=
\delta_c(t)+\dot{\varphi}_r(t)+\dot{\varphi}_b(t).
\label{eq:deltac_shift}
\end{align}
The phases of the two complex Rabi frequencies therefore disappear from the
off-diagonal couplings and reappear as shifts of the diagonal detunings.

Equations~\eqref{eq:Deltac_shift} and \eqref{eq:deltac_shift} show explicitly
that time-dependent laser phases are equivalent to time-dependent detunings.
The phase $\varphi_r(t)$ shifts the one-photon detuning $\Delta_c(t)$, whereas
$\varphi_r(t)+\varphi_b(t)$ shifts the two-photon detuning $\delta_c(t)$.

%\subsection{Inverse statement and physical interpretation}
%\label{app:control_phase_inverse}

Conversely, suppose one starts from real Rabi frequencies but allows the detunings
$\tilde{\Delta}_c(t)$ and $\tilde{\delta}_c(t)$ to vary in time. Introducing
reference detunings $\Delta_{c,0}(t)$ and $\delta_{c,0}(t)$, one may define phase
functions through
\begin{align}
\dot{\varphi}_r(t)
&=
\tilde{\Delta}_c(t)-\Delta_{c,0}(t),
\label{eq:phir_inverse}\\
\dot{\varphi}_b(t)
&=
\tilde{\delta}_c(t)-\delta_{c,0}(t)-\dot{\varphi}_r(t).
\label{eq:phib_inverse}
\end{align}
After integration, the phase-modulated Rabi frequencies
\begin{align}
\Omega_r(t)=|\Omega_r(t)|\,e^{i\varphi_r(t)},
\qquad
\Omega_b(t)=|\Omega_b(t)|\,e^{i\varphi_b(t)}
\end{align}
generate the same dynamics up to a time-dependent basis choice.  Hence,
optimizing $\tilde{\Delta}_c(t)$ and $\tilde{\delta}_c(t)$ is equivalent to
optimizing the phases $\varphi_r(t)$ and $\varphi_b(t)$.

Physically, a time-dependent phase is an instantaneous frequency chirp.  If a
laser field is written as
\begin{align}
E(t)\propto \cos\!\bigl[\omega t+\varphi(t)\bigr],
\end{align}
then its instantaneous frequency is
\begin{align}
\omega_{\mathrm{inst}}(t)=\omega+\dot{\varphi}(t).
\end{align}
Thus, changing the laser phase in time is operationally equivalent to changing
the detuning in time.  In the control-atom two-photon process, the first laser
phase controls the effective one-photon detuning, whereas the sum of both
phases controls the effective two-photon detuning.

In numerical optimal control, treating $\Delta_c(t)$ and $\delta_c(t)$ as
optimization variables is therefore mathematically equivalent to optimizing
the phases of the complex Rabi frequencies.  The two descriptions are related
by Eq.~\eqref{eq:Wc_phase_gauge} and parameterize the same driven dynamics.

\onecolumngrid
\section{Experimental parameter benchmarks}
\label{app:experimental_parameter_sets}

This appendix compiles the experimental benchmarks used in
Sec.~\ref{sec:experimental_mapping}. These records set the optical-control
bounds and calibrate the dissipative and technical-noise scales used in the
simulations. Table~\ref{tab:platform_laser_parameters} summarizes the optical
and geometrical parameters extracted from the source experiments.
Table~\ref{tab:platform_error_parameters} lists the associated noise scales
and the simulated $\mathrm{C}^1\mathrm{NOT}^{4}$ error entries. The first
benchmark column combines the EIT-CNOT data of Ref.~\cite{CNOT-neutral3} with
the single-atom imperfection analysis of Ref.~\cite{imperfection1}. The second
and third columns use high-fidelity Rydberg-gate benchmarks from
Refs.~\cite{experimental-setup2,evered_nonlocal_2026}, respectively.

\newcommand{\benchmarkentry}[2]{\shortstack{#1\\[-0.15ex]\scriptsize #2}}

\begin{table}[H]
\caption{
Optical and geometrical benchmark parameters used to define the control
window. In the first data column, paired detunings follow the order of
Refs.~\cite{CNOT-neutral3,imperfection1}. A dash denotes a quantity not
separately tabulated in the source record.
\label{tab:platform_laser_parameters}}
\centering
\footnotesize
\setlength{\tabcolsep}{3.2pt}
\renewcommand{\arraystretch}{1.13}
\begin{tabularx}{\textwidth}{@{}l*{3}{>{\centering\arraybackslash}X}@{}}
\toprule
Parameter & \cite{CNOT-neutral3,imperfection1} & \cite{experimental-setup2} & \cite{evered_nonlocal_2026} \\
\midrule
Species &
$^{133}\mathrm{Cs}$ &
$^{87}\mathrm{Rb}$ &
$^{87}\mathrm{Rb}$ \\
Rydberg state &
$81D_{5/2}$ &
$53S_{1/2}$ &
$53S_{1/2}$ \\
Spacing &
$6~\mu\mathrm{m}$ &
$\sim2~\mu\mathrm{m}$ &
$\sim1.73~\mu\mathrm{m}$ \\
Gate time &
$2~\mu\mathrm{s}$ &
$0.26~\mu\mathrm{s}$ &
-- \\
Wavelengths &
$852/509~\mathrm{nm}$ &
$420/1013~\mathrm{nm}$ &
$420/1015~\mathrm{nm}$ \\
Rabi frequencies &
\shortstack{$\Omega_r/2\pi=1.77~\mathrm{MHz}$\\
$\Omega_p^{\max}/2\pi=0.67~\mathrm{MHz}$\\
$\Omega_c/2\pi\sim40~\mathrm{MHz}$} &
\shortstack{$\Omega/2\pi=4.6~\mathrm{MHz}$\\
$\Omega_{420}/2\pi=237~\mathrm{MHz}$\\
$\Omega_{1013}/2\pi=303~\mathrm{MHz}$} &
\shortstack{$\Omega/2\pi\simeq17~\mathrm{MHz}$\\
$\Omega_{420}/2\pi=670~\mathrm{MHz}$\\
$\Omega_{1015}/2\pi=460~\mathrm{MHz}$} \\
One-photon detuning &
$\Delta/2\pi=870/740~\mathrm{MHz}$ &
$\Delta/2\pi=7.8~\mathrm{GHz}$ &
$\Delta/2\pi=7.8~\mathrm{GHz}$ \\
Two-photon detuning &
\shortstack{$\delta/2\pi=0.28(2)~\mathrm{MHz}$\\
$\Delta_c/2\pi\simeq1.8~\mathrm{MHz}$} &
$\delta_0=-0.9491\,\Omega$ &
$\delta_0=0.775\,\Omega$ \\
\bottomrule
\end{tabularx}
\end{table}

\begin{table}[H]
\caption{
Experimental noise scales and the corresponding simulated
$\mathrm{C}^1\mathrm{NOT}^{1}$ error entries. Each cell gives the benchmark
scale above the associated this-work infidelity $\epsilon_1=1-F_1$, expressed
in percent. The row labelled ``Interactions between target atoms'' instead gives the
interaction-induced $\epsilon_4=F_1^{\mathrm{ideal}}-F_4^{\mathrm{ideal}}$.
The final two rows list the full-noise fidelity $F_4$.
\label{tab:platform_error_parameters}}
\centering
\footnotesize
\setlength{\tabcolsep}{3.2pt}
\renewcommand{\arraystretch}{1.13}
\begin{tabularx}{\textwidth}{@{}l*{3}{>{\centering\arraybackslash}X}@{}}
\toprule
Channel or metric & \cite{CNOT-neutral3,imperfection1} & \cite{experimental-setup2} & \cite{evered_nonlocal_2026} \\
\midrule
Intermediate decay &
\benchmarkentry{$\Gamma_e/2\pi\simeq6~\mathrm{MHz}$}{$\epsilon_1=1.4098\%$} &
\benchmarkentry{$\tau_e=110~\mathrm{ns}$}{$\epsilon_1=0.4599\%$} &
\benchmarkentry{$\tau_e=110~\mathrm{ns}$}{$\epsilon_1=0.4599\%$} \\
Rydberg decay &
\benchmarkentry{--}{--} &
\benchmarkentry{$T_1=88~\mu\mathrm{s}$}{$\epsilon_1=0.1518\%$} &
\benchmarkentry{$T_1=72~\mu\mathrm{s}$}{$\epsilon_1=0.2209\%$} \\
Doppler dephasing &
\benchmarkentry{$k_{\mathrm{eff}}\sigma_v\simeq2\pi\times120~\mathrm{kHz}$}{$\epsilon_1=14.5795\%$} &
\benchmarkentry{$T_2^*=3~\mu\mathrm{s}$}{$\epsilon_1=7.4012\%$} &
\benchmarkentry{$T_2^*=4.7~\mu\mathrm{s}$}{$\epsilon_1=6.8939\%$} \\
Intensity noise &
\benchmarkentry{$<0.2\%$ rms}{$\epsilon_1=0.1563(825)\%$} &
\benchmarkentry{$0.25\%\!\rightarrow\!0.1\%$ drift}{$\epsilon_1=0.1139(819)\%$} &
\benchmarkentry{$\sim0.1\%$ stability}{$\epsilon_1=0.1092(694)\%$} \\
Phase noise &
\benchmarkentry{sub-kHz linewidth}{$\epsilon_1=0.0265(334)\%$} &
\benchmarkentry{$T_2^*=3~\mu\mathrm{s}$}{$\epsilon_1=0.0140(109)\%$} &
\benchmarkentry{$T_2^*=2.8$--$3.8~\mu\mathrm{s}$}{$\epsilon_1=0.0263(162)\%$} \\
Interactions between target atoms &
\benchmarkentry{$\eta=|C_6^{ct}/C_6^{tt}|\simeq40$}{$\epsilon_4=0.2624\%$} &
\benchmarkentry{$\eta=|C_6^{ct}/C_6^{tt}|\simeq82$}{$\epsilon_4=0.0331\%$} &
\benchmarkentry{$\eta=|C_6^{ct}/C_6^{tt}|\simeq82$}{$\epsilon_4=0.0331\%$} \\
\midrule
Total fidelity (unoptimized) &
$F_4=74.0148(4371)\%$ &
$F_4=91.1294(3335)\%$ &
$F_4=93.2806(2761)\%$ \\
Total fidelity (optimized) &
$F_4=90.0137(0916)\%$ &
$F_4=96.5453(0658)\%$ &
$F_4=97.8335(0617)\%$ \\
\bottomrule
\end{tabularx}
\end{table}

Following the error-budget convention used in high-fidelity Rydberg-gate
experiments, Table~\ref{tab:platform_error_parameters} distinguishes reported
noise strengths from infidelities obtained in the present simulations. For this
protocol, spontaneous emission is assigned to the intermediate-decay row,
because intermediate-state scattering dominates direct Rydberg decay on the
optimized gate timescale. 

\clearpage
\twocolumngrid
\bibliographystyle{apsrev4-1} % 选择参考文献格式
   \bibliography{references}   % 引用 references.bib 文件

\end{document}